\documentclass[12pt]{article}
\usepackage{amsmath}
\usepackage{graphicx}
\usepackage{enumerate}
\usepackage{natbib}
\usepackage{url} 

\usepackage{amsfonts}
\usepackage{amssymb}
\usepackage{xcolor}
\usepackage{hyperref}
\usepackage{float}

\providecommand{\U}[1]{\protect\rule{.1in}{.1in}}
{\tiny {\tiny {\large {\normalsize }}}}
\providecommand{\U}[1]{\protect\rule{.1in}{.1in}}
\newtheorem{theorem}{Theorem}

\newtheorem{algor)m}[theorem]{Algorithm}

\begin{document}

	\def\spacingset#1{\renewcommand{\baselinestretch}%
		{#1}\small\normalsize} \spacingset{1}


{
	\title{\textbf{Forecasting High Dimensional Time Series with Dynamic
			Dimension Reduction}}
	\author{Daniel Pe{\~n}a\thanks{%
			Partially supported by the Spanish Agencia Nacional de Evaluaci\'{o}n under
			grant PID2019-109196GB-I00 and by FUNCAS, Madrid} \hspace{.2cm} \\
		Department of Statistics, Universidad Carlos III de Madrid\\
		and \\
		V{\'i}ctor J. Yohai \thanks{%
			Partially supported by Grants 20020170100330BA from the Universidad de
			Buenos Aires and PICT 1302 from ANPYCT. We are grateful to Z. Gao and R. S.
			Tsay for providing the code to run their procedure}\\
		Department of Mathematics, Universidad de Buenos Aires}
	\maketitle
}  
{
	\bigskip
	\bigskip
	\bigskip
	\begin{center}
		{\LARGE\bf Forecasting High Dimensional Time Series with Dynamic Dimension 	Reduction}
	\end{center}
	\medskip
}  
	
	\bigskip
	
	\begin{abstract}
		
		{Many dimension reduction techniques have been developed for independent data, and most have also been extended to time series. However, these methods often fail to account for the dynamic dependencies both within and across series. In this work, we propose a general framework for forecasting high-dimensional time series that integrates dynamic dimension reduction with regularization techniques. The effectiveness of the proposed approach is illustrated through a simulated example and a forecasting application using an economic dataset.
			We show that several specific methods are encompassed within this framework, including Dynamic Principal Components and Reduced Rank Autoregressive Models. Furthermore, time-domain formulations of Dynamic Canonical Correlation and Dynamic Redundancy Analysis are introduced here for the first time as particular instances of the proposed methodology. All of these techniques are analyzed as special cases of a unified procedure, enabling a coherent derivation and interpretation across methods.}

	\end{abstract}
	
	\noindent%
	{\it Keywords:}  Dynamic Principal Components; Dynamic Canonical
	Correlation Analysis; Dynamic Redundancy analysis; Reduced rank
	autorregressive models; Regularization.
	\vfill
	
	\newpage
	\spacingset{1.9} 

\section{\protect\normalsize Introduction}

{\normalsize \label{intro} }

{\normalsize Building prediction rules for vector time series variables as a
function of a large set of other explanatory variables is an important
objective in Statistics, Machine Learning and Artificial Intelligence. In
this paper we will study linear prediction functions that can be used by
themselves, or as a starting point for more sophisticated non linear
approaches. For instance, instead of training a neural network directly on
the observed data it will be faster, and with better interpretation, to fit
first a linear model and use its residuals for searching for non linear
effects with a recurrent neural network (RNN). Then, both parts can be put
together in a RNN with linear part followed by a non linear structure, to
have a joint estimation of the forecasting rule. See, for instance, \cite{babu2014moving}.}

{\normalsize A standard procedure to build linear prediction rules is by the
multivariate regression model, where the parameters are estimated by least
squares. However, as it is well known, least squares estimation is not
efficient when the ratio between the number of explanatory variables and the
sample size is large, and several alternatives have been proposed for this
situation. Let us consider first the static case, in which we have a sample
of independent observations and we want to explain a response variable as a
function of a set of explanatory variables. }

{\normalsize One of the first approaches proposed for high dimensional
regression is variable selection, see \cite{Mallows73}, \cite%
{buhlmann2003boosting}, \cite{efron2004least}, \cite{wasserman2009high}, 
\cite{ing2011stepwise}, \cite{bertsimas2016best}, and \cite{hazimeh2020fast}.%
A second approach is dimension reduction, where the original explanatory
variables are replaced by a small set of linear combinations of them, that
summarize the relevant information. This method includes principal component
regression, partial least squares, redundancy analysis, projection pursuit
regression, reduced rank regression, factor analysis and structural equation
models, among other less popular methods. The third approach is
regularization (see \cite{lasso} and \cite{hastie2009elements}), where the
estimation function to be minimized is modified by introducing a
penalization for the size of the regression parameters, so that the effect
of the variables with small explanatory power is reduced, or even
eliminated. In this way, a kind of variable selection is made (\cite%
{huang2008adaptive}) }.

{\normalsize These three approaches are related (\cite{hastie2020best}). If
the relevant information is included in a subspace of small dimension, we
can take as explanatory variables a basis of this central subspace. In the
most interesting case in which the response is a vector of many correlated
variables, the central subspace includes the linear combinations that are
correlated with all the dependent variables,   reducing, some times enormously, the number
of parameter to be estimated. Regularization and variable selection achieve
a similar objective, by searching for a subspace formed by the most
important original variables. In this article we will use both dimension
reduction and regularization for forecasting vector time series response
variables. }

{\normalsize   Suppose we are interested in forecasting $y_{t+h}$, where $y_t$ is a single time series, 
using the values of a large vector time series of explanatory variables,  $\mathbf{x_t}$.
\cite{SW2002} proposed to assume that ($y_{t+h}, \mathbf{x_t}$) follow a factor model representation and estimate the factors using the leading principal components of the $\mathbf{x_t}$ variables. Then, the forecast of $y_{t+h} $ is made by using these factors and the lag values of the $y_t$ as explanatory variables and estimating the factor loadings and autorregressive coefficients by least squares. They showed that,  under general conditions, these
estimates are shown to be consistent.  As some of the leading principal
components may not be the most useful for forecasting, several improvements
of this procedure have been proposed in order to select the best factors for
prediction. For instance, \cite{bai2009boosting} applied boosting and
thresholding rules to select them and \cite{stock2012generalized} used
regularization to down-weight non important factors. When instead of
a single time series we want to forecast a set of $q$ dependent time series and, assuming that the set of explanatory variable is common to all of
them, use the same factors, $\widehat{\mathbf{f}}_{t},$  for each of the
series. Also, specific factors for each of the series can be computed from different sets
of explanatory variables. Other approaches are based on other dimension
reduction methods, as for instance, sliced inverse regression, see 
\cite{barbarino2024forecast}, \cite{matilainen2019sliced}, partial least
squares, see \cite{fuentes2015sparse}, and other related methods, see \cite%
{nordhausen2021dimension} and \cite{fan2017sufficient}. }

{\normalsize \cite{huang2022scaled} proposed to scale the PC by regressing
the variable to be predicted on each of the $p$ predictors, 
\begin{equation}
\widehat{y}_{t+h}=\widehat{\alpha}_{i}+\widehat{\gamma}_{i}x_{it} =\widehat{%
\alpha}_{i}+\widehat{x}_{it},  \label{sdPCA}
\end{equation}
and instead of computing the principal components of the original variables
use those of the scaled variables $\widehat{x}_{it}=\widehat{\gamma}%
_{i}x_{it}$. The method is called scaled PC, as the explanatory variables
are scaled by their predictive power. Then, calling $\widehat{\mathbf{f}}%
_{t}^{sc}$ to the principal components corresponding to the $s$ largest
eigenvalues of the scaled variables $\widehat{x}_{it},$ the forecast is made
  using $\widehat{\mathbf{f}}_{t}^{sc}$ as
factors. They showed that this procedure improves the forecasting
performance with respect to using the standard principal components. Note
that if we want to forecast a vector of time series this method always
computes specific factors for each of the series. An extension of this idea
was proposed by \cite{gao2024supervised} as the Supervised
Dynamic Principal Component Analysis (sdPCA). First, for $i=1,...,p,$ they
run the regressions including also lags in the relationship of the
explanatory variable considered 
\begin{equation}
\widehat{y}_{t+h}=\widehat{\alpha}_{i}+\widehat{\gamma}_{i,0}x_{t,i} +...+%
\widehat{\gamma}_{i,q_{i}-1}x_{t-q_{i}+1,i}=\widehat{\alpha} _{i}+\widehat{x}%
_{t,q_{i},i},  \label{GT}
\end{equation}
and, second, the main principal components of the variables $\widehat{x}%
_{t,q_{i},i}$ are obtained. The forecast is made by using the regression of
the variables ${y}_{t+h}$ on the selected principal components, ${p}_{i,t}$.
They showed the good practical performance of his proposal compared to some
previous methods. }

{\normalsize An approach oriented to forecasting the common part of a vector
of time series was proposed by Forni et al. (2005). They assume that the
vector $\mathbf{y}_{t}$ follows a generalized dynamic model given by 
\begin{equation*}
\mathbf{y}_{t}=\boldsymbol{\chi}_{t}+\mathbf{\xi}_{t}=\mathbf{C}(B)\mathbf{f}%
_{t}+\mathbf{\xi}_{t}=\mathbf{PF}_{t}+\mathbf{\xi}_{t},
\end{equation*}
where $\mathbf{y}_{t}$ is decomposed in two orthogonal components, a common
component $\boldsymbol{\chi}_{t}$ and an idiosyncratic component $\mathbf{%
\xi }_{t}.$ The common component is the product of a loading matrix, $\mathbf{%
C}(B)=\mathbf{C}_{0}+\mathbf{C}_{1}B+...+\mathbf{C}_{s}B$, and a vector of $d$
factors, $\mathbf{f}_{t}$, and it can also be written as a static factor
model with $r=d(s+1)$ factors, $\mathbf{F}_{t}$, and loading matrix,  $\mathbf{%
P}$,  of dimensions $q\times r.$ These authors forecast the common part with $%
\widehat{\mathbf{P}}\widehat{\mathbf{F}}_{t}$, estimated by using the main
eigenvectors of the spectral matrix. An alternative approach that also takes
into account the information in the lags of the series was proposed by \cite%
{pena2019} with the one-sided dynamic principal components (ODPC). They are
defined as linear combinations of the observations based on a one-sided
filter of past and present observations. The forecast with ODPC requires to
model the components to predict their future values, whereas in the
procedure proposed in this article this is not necessary. Instead, their
loadings in the prediction equation depend on the forecasting horizon. ODPC
will be briefly revised in Section \ref{DPCF} as a particular simple case of the
core component method we propose in this article. }

{\normalsize Note that when the vector, $\mathbf{y}_{t}$, to be forecast is
composed of independent series there is no possible advantage in considering
jointly the forecast of all of them. However, when the cross-correlations
among the series are strong we can estimate better the common part by using
all the \ $\mathbf{y}_{t}$ series as additional explanatory variables,
instead of considering each series individually. In this way we may find
other factors useful for forecasting that are not included in the set $%
\mathbf{x}_{t}$ of explanatory variables. }

{\normalsize In this work we propose a procedure to forecast
a $q$- dimensional vector time series, $\mathbf{y}_{t}$, by using a large
set of explanatory time series variables $\mathbf{z}_{t}$. This set may
include some and even all the components of $\mathbf{y}_{t}$. The procedure
first builds an enlarged set of explanatory variables, $\mathbf{x}_{t}$, of
dimension $p$, including the series in $\mathbf{z}_{t}$ and also a given
number of their past values. Then, the forecast is a linear function,  
which depends on the 
forecast horizon,  of a
small number of components, built as linear combinations of the $\mathbf{x}%
_{t}$ variables, that will be called core components. These core components
and some of their lags are used in the forecast equation as explanatory
variables. This prediction can be thought as obtained by a dynamic
reduced-rank linear model that approximates the unknown relationship between 
$\mathbf{z}$ and $\mathbf{y}$. }

Our procedure has three features that are not usually found
jointly in previous dimension reduction approaches for forecasting vector
time series. First, it does not assume any model for the series involved
 and do not assume the same linear form of the forecast equation for all horizons.
Second, it allows for a different number of lags of the explanatory variables
in building each of the core components and, also, a different number of
lags of each the core component in the forecasting equation. Third, the
structure of the factors and all the coefficients required in the
forecasting equation are estimated jointly, minimizing a prediction loss
function, instead of first finding with some procedure the factors and second 
estimating their effect on the forecasting equation. 

{\normalsize There are two main contributions in this article. The first one
is to present a dimension reduction forecasting procedure for high
dimensional set of related time series, using a small number of linear
combinations of the predictor variables. We also introduce regularization,
assuming that the coefficients of these linear combinations are sparse. The
objective function is the prediction mean squared error with a lasso type
penalization. We show that this procedure, that will be called the core
components forecasting (CCF) procedure, works well in applications and that
the regularization has the additional advantage of making the core
components easier to interpret. The second contribution is to show that
under this forecasting approach dynamic versions of principal components,
canonical correlation and redundancy analysis appear as particular cases
defined as linear combinations of the present and lagged valued of the $%
\mathbf{z}_{t}$ 's optimizing their prediction power. Dynamic principal
components and redundancy analysis appear when minimizing the sum of the
mean squared prediction errors of each variable, whereas dynamic canonical
correlation is  when minimizing the determinant of the matrix of
prediction errors. }

{\normalsize The rest of this article is organized as follows. Section 2
presents the general approach to the forecasting problem considered and its
estimation under two different estimation criteria.  Section 3 introduces the proposed
core components forecasting(CCF) procedure, based on a square loss function
with sparse estimation by lasso penalization. Section 4 describes  some asymptotic
properties: for stationary and ergodic time series the non-regularized parameter estimators
converge to their population analogues. Section 5 indicates how to
select the  forecasting rule, that is,  how to choose the number of lags  and the number of components 
used in the procedure by means  of cross validation. Section 6 shows that dynamic
principal components, dynamic canonical analysis and dynamic redundancy
analysis are particular cases of our general formulation presented in
Section 2. Section 7 illustrates the performance of the core components
forecasting procedure in an Monte Carlo Simulation example and Section 8
in the analysis of a real data set. Finally, Section 9 includes some
concluding remarks. The Appendix contains the proofs of the theorems. }

\section{\protect\normalsize A General Dimension Reduction Approach to
Forecast large sets of Time Series}
{\normalsize \label{ccfp} }

\subsection{\protect\normalsize Definition of the Core Components}
  
  {\normalsize Let $\mathbf{y}_{t}=(y_{t,1},...,y_{t,q})^{\prime },$ $1\leq
t\leq T$, be a multiple time series of dimension $q$. We are interested in
forecasting $\mathbf{y}_{T+h}=(y_{T+h,1},...,y_{T+h,q})^{\prime }$, for $%
h\geq 0$, using a large set of time series $\mathbf{z}%
_{t}=(z_{t,1},...,z_{t,m})^{\prime }$. This set may also include as
explanatory variables of $\mathbf{y}_{T+h}$ some, or all, of the components
of $\mathbf{y}_{t}$. The forecast function will use these series and $c_{1}$
of their lags, where }$c_{1}$ is a non negative integer number to be chosen in the procedure as to be explained later, {\normalsize %
\ so that the set of of explanatory variables is $\mathbf{x}_{t}=(\mathbf{z}^\prime%
_{t},\mathbf{z}^\prime_{t-1} ,...,\mathbf{z}^\prime_{t-c_1})^\prime$, $(c_{1}+1)\leq t\leq T$ and
includes $p=m(c_{1}+1)$ variables. We will not make any assumptions about
the probability distributions of the series $(\mathbf{z}_{t},\mathbf{y}_{t})$.
 In particular, we do not assume that they are stationary. }

{\normalsize When $h=0$, the procedure we propose instead of forecasting
future values explains, or  reconstructs, the observed data of the series $%
\mathbf{y}_{t}$ using a smaller set of variables that includes most of the
information in the data set. This reconstruction problem is applied in
signal processing and information networks. When  $h>0$, the problem of forecasting is
of more general interest in many areas, and is the objective of this article.}

{\normalsize Assuming that $p$ is large, we will be looking for a small set of core
components, which are linear combinations of $\mathbf{x}_{t}$, to be used to
forecast $\mathbf{y}_{T+h}$. We build these components in an iterative way.
We assume, without loss of generality, that the variables $\mathbf{x}_{t}$
and $\mathbf{y}_{t}$ have zero means. The first core component will be a
time series of the form, 
\begin{equation}
f_{t}(\boldsymbol{\beta }_{h})=\mathbf{x}_{t}^{\prime }\boldsymbol{\beta }
_{h},\ c_1+1\leq t\leq T  \label{lcomp}.
\end{equation}%
Let }$k_{1}$ be another non negative integer number to be chosen. Then,  {\normalsize the
univariate time series, $f_{t}$,  and $k_{1}$ of their lags, will be used to
predict the vector $\mathbf{y}_{T+h}$ in a linear way. Let 
\begin{equation}
\mathbf{f}_{t}\ (\boldsymbol{\beta }_{h})=(f_{t}(\boldsymbol{\beta }%
_{h}),f_{t-1}(\boldsymbol{\beta }_{h}),...,f_{t-k_{1}}(\boldsymbol{\beta } %
_{h}))^{\prime }=\mathcal{X}_{t}\boldsymbol{\beta }_{h},  \label{1compk}
\end{equation}%
where $\mathcal{X}_{t}$ is a $(k+1)\times p$ matrix with rows $\mathbf{x}%
_{t},\mathbf{x}_{t-1},...,\mathbf{x}_{t-k}$ and let $\boldsymbol{\Gamma }%
_{h} $ {be a} $q\times (k_{1}+1)$  matrix. Then, the prediction is defined by } 
{\normalsize 
\begin{equation}
\widetilde{\mathbf{y}}_{t+h|t}(\boldsymbol{\beta }_{h}\boldsymbol{,\Gamma }%
_{h}\boldsymbol{)}\ =\boldsymbol{\Gamma }_{h}\mathbf{f}_{t}(\boldsymbol{%
\beta }_{h}\boldsymbol{)=\Gamma }_{h}\mathcal{X}_{t}\boldsymbol{\beta }_{h},%
\text{ }c_{1}+k_{1}+1\leq t\leq T, 
\end{equation}%
}
and the prediction errors when the parameters $\boldsymbol{\beta }_{h}\boldsymbol{%
\ }$\ and $\boldsymbol{\Gamma }_{h}$ are used, are  {\normalsize 
\begin{equation*}
\mathbf{e}_{t+h|t}^{(1)}(\boldsymbol{\beta }_{h}\boldsymbol{,\Gamma 
}_{h}\boldsymbol{)}=\mathbf{y}_{t+h}-\widetilde{\mathbf{y}}_{t+h|t}(%
\boldsymbol{\beta }_{h}\boldsymbol{,\Gamma }_{h}\boldsymbol{)},\text{ }%
c_{1}+k_{1}+1\leq t\leq T-h. 
\end{equation*}
}

The values of $(\boldsymbol{\beta }_{h}\boldsymbol{,\Gamma }_{h}\boldsymbol{)%
}$ given the values of $c_1$ and $k_1$ should be estimated so that \ the residuals $\mathbf{e}_{t+h|t}^{%
(1)}(\boldsymbol{\beta }_{h}\boldsymbol{,\Gamma }_{h}\boldsymbol{)},$ $%
c_{1}+k_{1}+1\leq t\leq T$ are as small as possible. \ Different criteria for estimating 
these parameters are proposed in the next subsection. Let $(\widehat{%
\boldsymbol{\beta }}_{h}^{(1)}\boldsymbol{,}\widehat{\boldsymbol{\Gamma }}%
_{h}^{(1)}\boldsymbol{)}$ be the estimators of $(\boldsymbol{\beta }_{h}%
\boldsymbol{,\Gamma }_{h}\boldsymbol{),}$ then, the first core component is
the time series $f_{t}(\widehat{\boldsymbol{\beta }}_{h}^{(1)}),$ $%
c_{1}+1\leq t\leq T$, \ and the {\normalsize forecasted values using these
parameter estimates are given by
\begin{equation}
\widehat{\mathbf{y}}_{t+h|t}=\widehat{\boldsymbol{\Gamma }}_{h}^{(1)}\mathbf{%
f}_{t}(\widehat{\boldsymbol{\beta }}_{h}^{(1)}\boldsymbol{)=}%
\widehat{\boldsymbol{\Gamma }}_{h}^{(1)}\mathcal{X}_{t}\widehat{\boldsymbol{%
\beta }}_{h}^{(1)},    \text{ }c_{1}+k_{1}+1\leq t\leq T, 
\label{pred1}
\end{equation}%
with prediction errors,  
\begin{equation*}
	\widehat{\mathbf{e}}_{t+h|t}^{(1)}=\mathbf{y}_{t+h}-\widehat{%
		\mathbf{y}}_{t+h|t},
\end{equation*}%
and the prediction of $\mathbf{y}_{T+h}${\normalsize \ using only the first
core component } is 
\begin{equation*}
\widehat{\mathbf{y}}_{T+h|T}^{(1)}=\widehat{\boldsymbol{\Gamma }}_{h}^{(1)}%
\mathbf{f}_{T}(\widehat{\boldsymbol{\beta }}_{h}^{(1)}\boldsymbol{)}.
\end{equation*}%
To improve this prediction we may use more core components. The second core
component is defined the same way as the first one, but taking as variables
to predict the residuals }$\widehat{\mathbf{e}}_{t+h|t}^{(1)}$%
{\normalsize \ instead of }$\mathbf{y}_{t+h}${\normalsize . Higher order
components are defined similarly. More precisely, suppose that we have
already computed }$i${\normalsize \ core components, using the given values  }$%
c_{j}$ and $k_{j}$ {\normalsize for the component }${\normalsize j \leq i,}$ 
{\normalsize \ $\ $and let \ }$\widehat{\mathbf{e}}_{t+h}^{(i)},$ $%
d_{i}+1\leq t\leq T-h,$ where $d_{i}=\max_{1\leq j\leq i}(c_{j}+k_{j}),$ the
residuals after using these $\ i$ components. Then,$\ (\widehat{\boldsymbol{%
\beta }}_{h}^{(i+1)},\widehat{\boldsymbol{\Gamma }}_{h}^{(i+1)})$ 
{\normalsize \ are defined as in the case }${\normalsize i=1}$, but with \ $%
\mathbf{y}_{t},1\leq t\leq $ $T$, replaced by $\widehat{\mathbf{e}}%
_{t+h|t}^{(i)}$, $d_{i}+1\leq t\leq T-h$. Besides, the{\normalsize \ }$%
{\normalsize (}i+1)$-th \ core component is the time series {\normalsize \ }$%
f_{t}${\normalsize $(\widehat{\boldsymbol{\beta }}_{h}^{(i+1)})$, }$%
d_{i+1}\leq t\leq T-h$, {\normalsize \ }and the new residuals are {\normalsize 
\begin{equation}
\widehat{\mathbf{e}}_{t+h|T}^{\mathbf{(}i+1\mathbf{)}}=\widehat{\mathbf{e}}%
_{t+h}^{(i)}-\widehat{\boldsymbol{\Gamma }}_{h}^{(i)}\mathbf{f}_{T}^{\ \ \ }(%
\widehat{\boldsymbol{\beta }}_{h}^{(i)}),\      c_{i+1}+k_{i+1}+1\leq
t\leq T.   \label{residd}
\end{equation}%
}

{\normalsize \ Finally, \ the prediction of $\mathbf{y}_{T+h},$ }when $s$ 
{\normalsize components are used is,%
\begin{equation}
\widehat{\mathbf{y}}_{T+h|T}^{(s)}=\sum_{i=1}^{s}\widehat{\boldsymbol{\Gamma 
}}_{h}^{(i)}\mathbf{f}_{T}^{\ \ \ }(\widehat{\boldsymbol{\beta }}_{h}^{(i)}).
\label{predfor}
\end{equation}%

{\normalsize Our approach is different from the procedure proposed by \cite%
{Reinsel1998} for reduced rank vector autoregressive processes (RRVAR). First, we 
include explanatory variables, that were not considered in the RRVAR. Second, 
in the case of AR($p$)\ processes without explanatory
variables they generate  forecasts 
by building models for the
time series component, $\mathbf{f}_{t}(\boldsymbol{\beta _{0})}$, and also
for the residuals $\mathbf{e}_{t}(\boldsymbol{\beta _{0},\Gamma }_{0}%
\boldsymbol{)}$, and calling $\widehat{\mathbf{f}}_{t+h|t}(\boldsymbol{\beta
_{0})}$ and $\widehat{\boldsymbol{\mathbf{e}}}\boldsymbol{_{t+h|t}(%
\boldsymbol{\beta _{0},\Gamma }_{0}\boldsymbol{)}}$ to the forecast
generated by these models, the $h$ time ahead forecast would be }
{\normalsize 
\begin{equation}
\tilde{\mathbf{y}}_{t+h|t}=\boldsymbol{\Gamma_{0}}\widehat{\mathbf{f}}
_{t+h|t}(\boldsymbol{\beta_{0})+\widehat{\mathbf{e}}_{t+h|t}(\boldsymbol{\
\beta _{0},\Gamma_{0})}},  \label{pred2}
\end{equation}
that will usually be quite different from (\ref{pred1}),   which does not assume any model and have coefficients which  depend on the horizon.  Third, we do not assume stationarity or impose any specific model on the time series.

It should be noticed that, the core components may imply a large
reduction in the number of parameters with respect to using multivariate
regression to compute a prediction $\widetilde{\mathbf{y}}_{T+h}$ as a
function of the $p$ components of $\mathbf{x}_{t}$. 
The regression parameter matrix  for multivariate regression will require $%
qp$ parameters, whereas each core component (see equation (\ref{pred1}))
includes $p+(k+1)q$ parameters and, if $s$ core components are fitted, the
total number of parameters in the core component procedure, assuming to simplify that all the components are computed with the same values of $c$ and $k$, will be $sp+sq(k+1)$. The ratio between the number of 
parameters with core components and multivariate regression will be $s/q + s(k+1)/m(c+1)$,
that will be usually small, as $m$ and $q$ are expected to be  much larger than $s$. Also, note that the reduction in the number of parameters depend on the sizes of the lags $k$ and $c$. Increasing the lag $k$ one unit will add $q$ parameteres whereas increasing the lag $c$ one unit will add $m$ parameters. 

  As shown by \cite{SW2002}, often the forecast of and individual time series can be improved
by adding to the information obtained from a set of explanatory variables a specific term corresponding to the lags of the series to be forecasted. In our case, once the forecast {\normalsize (\ref{predfor})
	are computed and the past forecast errors }$e_{j,t+h|t}=y_{j,t+h}-\widehat{y}%
_{j,t+h|t}$ 
are obtained,  this will imply fitting an autoregressive model to these errors and add this representation  to the prediction. Thus,  the forecast of the
component $y_{j,t},$ for $1\leq j\leq q$,  will be given by 
\begin{equation*}
	\widehat{y}_{T+h|T,j}^{(s)}=\sum_{i=1}^{s}\widehat{\boldsymbol{\gamma }}%
	_{h,j}^{(i)}\mathbf{f}_{T}^{(i)}+\sum_{r=1}^{p_{j}}\widehat{\phi }_{h,j,r}y_{(T-r),j},
\end{equation*}%
where the row vector $\widehat{\boldsymbol{\gamma }}_{h,j}^{(i)}$ is the $j$-th
row of the matrix $\widehat{\boldsymbol{\Gamma }}_{h}^{(i)}$ and the
coefficients $\widehat{\phi }_{h,j,r}$ are estimated by least squares from the past
forecast errors $\widehat{e}_{t+h,j|t}$. 

 Note that, since the residuals given in (\ref{residd}) depend on $h$ the estimators  of the parameters    $\widehat{\boldsymbol{\Gamma}}_{h}^{(i)},$
and $\widehat{\boldsymbol{\beta}}_{h}^{(i)}$ are  going to depend on $h$  too,   and consequently, they can adapt 
to a non linear evolution of the future values of the time series.
Nonetheless, by the sake of simplicity, from now on, we will assume a given forecast 
horizon, $h$,  and eliminate the subscript $h$ from the parameters.
  
  In the next subsection we will propose different criteria  for estimating $\boldsymbol{\beta}$ and  $\boldsymbol{\Gamma}$.   The choice of   the number of core components $s$, and  the lags $c_{i}$ and $k_{i}$ used for each component,    $1\leq
i\leq s$, is discussed in Section \ref{cross}. Several possible
problems covered by this setup will be analyzed in Section \ref{PCCCFA}. 

 \subsection{\protect\normalsize Criteria for estimating the Core Components} \label{criteria}
{\normalsize When the dimension of $\mathbf{x}_{t}$, $p$, is large compared
with the number of observations, $T$, the amount of parameters required to
define the components also increases. Then, as it occurs in other
statistical problems, the classical estimation procedures may not be
reliable. A good solution is to regularize the estimation by adding to the
loss functions a penalization. We choose the lasso form, although other
alternatives can be used, (see for instance \cite{hastie2015statistical}). }

{\normalsize The loss function to estimate the prediction rule can be based
on making the forecast errors as small as possible. Two main criteria with
this objective have been proposed. The first one is least squares,
minimizing the average over time of the Euclidean norm of the forecast error
vectors, that is, the 	 norm of the matrix of forecast errors. Thus,
adding the penalization to this criterion the estimates will be computed as }
{\normalsize 
\begin{equation}
(\widehat{\boldsymbol{\beta }}\boldsymbol{,}\widehat{\boldsymbol{\Gamma }
}\boldsymbol{)=}\arg \min_{\boldsymbol{\boldsymbol{\beta ,\Gamma }}%
}RG_{1} \boldsymbol{(\boldsymbol{\beta ,\Gamma )}} ,  \label{mRG1}
\end{equation}
where 
\begin{equation}
RG_{1}(\boldsymbol{\beta ,\Gamma})=G_{1}\boldsymbol{( \boldsymbol{\beta
,\Gamma )+}}\lambda \frac{   \boldsymbol{||\beta ||}_{1}}{|| \boldsymbol{\beta
||}_{2}},  \label{RG1}
\end{equation}
and 
\begin{eqnarray}
G_1(\boldsymbol{\beta }, \boldsymbol{\Gamma }) &=& \frac{1}{(T-h-c-k)}%
\sum_{t=k+c+1}^{T-h}\boldsymbol{|| \mathbf{e}}_{t+h|t}\boldsymbol{(%
\boldsymbol{\beta ,\Gamma )}||}_{2}^{2}  \notag \\
&=& \frac{1}{(T-h-c-k)}\sum_{t=k+c+1}^{T-h} \left \vert \left\vert \mathbf{y}%
_{t+h}-\boldsymbol{\Gamma }\mathcal{X}_{t} \boldsymbol{\beta }\right\vert
\right\vert _{2}^{2} ,  \label{G1}
\end{eqnarray}
and $||\mathbf{a}||_d$ denote the L$^d$ norm of a vector $\mathbf{a}$. }

{\normalsize The second criterion minimizes the determinant of the sample
covariance matrix of the residuals and is often used with normal data.
Adding a lasso penalization to this criterion the estimates are obtained by 
\begin{equation}
(\widehat{\boldsymbol{\beta }}\boldsymbol{,}\widehat{\boldsymbol{\Gamma }
}\boldsymbol{)=}\arg \min_{\boldsymbol{\boldsymbol{\beta ,\Gamma }}%
}RG_{2} \boldsymbol{(\boldsymbol{\beta ,\Gamma )}} ,  \label{mRG2}
\end{equation}
where, 
\begin{equation}
RG_{2}(\boldsymbol{\beta ,\Gamma})=G_{2}\boldsymbol{( \boldsymbol{\beta
,\Gamma )+}}\lambda \frac{||\boldsymbol{\beta ||}_{1}}{|| \boldsymbol{\beta
||}_{2}},  \label{RG2}
\end{equation}
and, 
\begin{equation}
G_2(\boldsymbol{\beta }, \boldsymbol{\Gamma }) = \left\vert \frac {1 }{%
(T-h-c-k)}\sum_{t=k+c+1}^{T-h}\boldsymbol{\mathbf{e}}_{t+h|t} \boldsymbol{( 
\boldsymbol{\beta,\Gamma)}}\mathbf{e}_{t+h|t}(\boldsymbol{\beta ,\Gamma)}
^{\prime}\right\vert ,  \label{G2}
\end{equation}
where $|\mathbf{A}|$ denotes determinant of $\mathbf{A}$. As it is well
known, if instead of measuring the size of the expected residuals covariance
matrix by the determinant we use the trace of the matrix, we go back to
least squares. }

{\normalsize As for any scalar $\gamma$, and $i=1,2$, $G_i(\gamma\boldsymbol{%
\beta },\boldsymbol{\Gamma} /\gamma)=G_i(\boldsymbol{\beta},\boldsymbol{%
\Gamma} )$, and thus $RG_i(\gamma\boldsymbol{\beta},\boldsymbol{\Gamma}
/\gamma)=RG_i(\boldsymbol{\beta},\boldsymbol{\Gamma)}$, we can assume in (%
\ref{mRG1}) and (\ref{mRG2}) that $||\boldsymbol{\beta }||_{2}=1. $ }

{\normalsize Note that in the multivariate regression model estimating the
regression parameters by minimizing the trace or the determinant of the
residual covariance matrix are the same. Besides, under normal errors, these
estimators coincide with the maximum likelihood estimator (see, for
instance, \cite{seber2009multivariate}). On the other hand, for the reduced
rank regression model, both criteria are different and when the regression
errors are normal, the maximum likelihood estimator is the one minimizing
the determinant of the residual matrix (see \cite{stoica1996}).
Moreover, it is also the criterion that leads to canonical correlation
analysis when you want to predict a vector $\mathbf{y}$ as a linear
combination of a vector $\mathbf{x}$ (see \cite{yohai1980canonical}). }

{\normalsize However, for forecasting the criterion based on $G_1$ seems preferable.
 An example  where  this criterion  seems to work better occurs  when there exists one components of $\mathbf{y%
}_{t+h}$ that can be predicted by $\boldsymbol{\beta}_{0}^{\prime}\mathbf{x}$
with prediction error close to zero. In that case, defining $\widehat{%
\boldsymbol{\ \beta}}$ $=\boldsymbol{\beta}_{0}$ will have the row and
column corresponding to this component in the errors covariance\ matrix
close to zero, and, therefore, the determinant of this matrix will be close to
zero too. Then, this will imply that $\widehat{\boldsymbol{\beta}}$ will be
close to $\boldsymbol{\beta }_{0} $ even if this coefficient is a very bad
predictor of the other components of $\mathbf{y}_{t+h}.$ This situation
cannot occur with loss function (\ref{G1}), and for that reason it is chosen to be used
in our proposed forecasting procedure. }

\label{EEQ}

\section{\protect\normalsize Computing the Core Components} {\normalsize \label{CCE} }
To compute the core components for prediction we will use the regularized
least squares loss function (\ref{RG1}). However, since the estimation of
the components that minimized the determinant loss function (\ref{RG2}) can
be useful for other applications and can be carried out in a similar way, we
will explain the minimization of both functions. With this objective, will
refer to a general loss function $RG_{i}(\boldsymbol{\beta },\boldsymbol{%
\Gamma })$, for $i=1,2$. given by (\ref{G1}) and (\ref{G2}). 

Note that given $\boldsymbol{\beta }$, the value of  $\boldsymbol{\Gamma}$ that minimizes 
$RG_{i}(\boldsymbol{\beta },\boldsymbol{\Gamma })$  is the coefficient matrix of a multivariate regression estimator where
the $(T-c-k-h)\times (k+1)$ regressor matrix $\mathbf{F}(\boldsymbol{\beta})$ has as $t$-th row   $\mathbf{f}_t^{\prime}(\boldsymbol{\beta})$,  and the $( T-h-c-k)\times q$ outcome matrix $\mathbf{Y}$ has as $t$-th row $\mathbf{y}_{t+h+c+k}$. Then, since in the case that the loss function is $RG_1$ we have to minimize the sum of the squares of the residuals, the value of                               $\boldsymbol{\Gamma }$ is obtained applying   least squares at each component of $\mathbf{y}_{t+h+c+k}$. Then $\boldsymbol{\beta}$ is given by $\mathcal{G}(\boldsymbol{\beta}) $, where 
\begin{equation}
\mathcal{G}(\boldsymbol{\beta}) 
= \mathbf{Y}^{\prime}\mathbf{F}(\boldsymbol{\beta})(\mathbf{F}(\boldsymbol{\beta})^{\prime}\mathbf{F}(\boldsymbol{\beta}))^{-1}. \label{Gammaa}
\end{equation} 
 When the loss function is $RG_2$ the value of $\boldsymbol{\Gamma}$ should be obtained by minimizing the determinant of the residual matrix $\mathbf{Y}-\mathbf{F}\boldsymbol{\Gamma}$. But, as mentioned in  subsection  \ref{criteria}, for a  multivariate regression model this criteria coincides with the  minimization of the sum of all the squared residuals.
Therefore, given $\boldsymbol{\beta}$, the value of $\boldsymbol{\Gamma}$ is also given by \ref{Gammaa}.

    
Put, for $i=1,2$,
\begin{equation*}
G_{i}^{\ast }(\boldsymbol{\beta })=G_{i}(\boldsymbol{\beta },\mathcal{%
G}(\boldsymbol{\beta})),
\end{equation*}
where  $ \mathcal{G}(\boldsymbol{\beta })$ is defined by (\ref{Gammaa}). We are
going to consider the regularized loss function  
\begin{equation*}
RG_{i}^{\ast }(\boldsymbol{\beta })=G_{i}^{\ast }(\boldsymbol{\beta }%
)+\lambda \frac{{\normalsize ||}\boldsymbol{\beta ||}_{1}}{{\normalsize ||}%
\boldsymbol{\beta ||}_{2}},
\end{equation*}
and the first core component estimator of $\boldsymbol{\beta }$ can also be  defined
by 
\begin{equation*}
\widehat{{\boldsymbol{\beta }}}=\arg \min_{  
||\boldsymbol{\beta }||_2=1}RG_{i}^{\ast }(\boldsymbol{\beta }), \label{beta2}
\end{equation*}%
and 
\begin{equation}
\widehat{{\normalsize \boldsymbol{\Gamma }}} 
= \mathcal{G}(\widehat{{\boldsymbol{\beta }}}). \label{gama2}                                           
\end{equation}

As previously indicated, without loss of generality we can take $||\widehat{ 
\boldsymbol{\beta }}||_{2}=1.$ To estimate $\widehat{{\normalsize 
\boldsymbol{\beta }}}$ and $\widehat{{\normalsize \boldsymbol{\Gamma }}}%
$ {\normalsize we will use an iterative algorithm based on proximal
gradient descent (see \cite{Parikh2013}). To this end, we start by finding
the gradient of $RG_{i}^{\ast }(\boldsymbol{\beta })$ with respect to $%
\boldsymbol{\ \beta }$ when $||\boldsymbol{\beta }||_{2}=1$. A simple
calculus shows that if $\boldsymbol{\beta }=(\beta _{1},...\beta
_{p})^{\prime }$ and $\left\vert \left\vert \boldsymbol{\beta }\right\vert
\right\vert _{2}=1$, then 
\begin{equation}
\mathbf{g}_{i}^{\ast }(\boldsymbol{\beta )}=\frac{\partial RG_{i}^{\ast }(\boldsymbol{\beta })}{\partial \boldsymbol{%
\beta }}=\frac{\partial G_{i}^{\ast }(\boldsymbol{\beta })}{\partial 
\boldsymbol{\beta }}-\lambda ||\boldsymbol{\beta }||_{1}\boldsymbol{\beta }%
+\lambda \boldsymbol{I({\beta }}\neq 0),  \label{gr}
\end{equation}%
where $\boldsymbol{I({\beta }} \neq 0)$ is a vector with $i$-th component$%
I(\beta_{i} \neq 0)=\text{sgn}(\beta _{i})$. Calling $\mathbf{g}_{i}(\boldsymbol{\beta )}$ to the continuous
part of (\ref{gr}), that is, 
\begin{equation}
\mathbf{g}_{i}(\boldsymbol{\beta )}=\frac{\partial G_{i}^{\ast }(\boldsymbol{%
\beta })}{\partial \boldsymbol{\beta }}\boldsymbol{-}\lambda ||\boldsymbol{%
\beta }||_{1}\boldsymbol{\beta},  \label{grad}
\end{equation}
then, $\mathbf{g}_{i}^{\ast }(\boldsymbol{\beta )} =\mathbf{g}_{i}(\boldsymbol{\beta} )+\lambda \mathbf{I}(\boldsymbol{%
\beta}\neq 0)$. This implies that   $\mathbf{g}_{i}\boldsymbol{(%
\boldsymbol{\beta )}}$ is equal to   the  gradient when all the components of the $%
\boldsymbol{\beta}$ vector are different from zero. We can control the case that  some components of $\boldsymbol{\beta}$ are zero 
by using the soft thresholding function (see \cite{Parikh2013}). This
function is defined for a scalar, $v$ by 
\begin{equation}
S_\gamma({v})=\left\{ 
\begin{array}{cc}
v-\gamma & \text{if }v>\gamma \\ 
0 & |v|\leq \gamma \\ 
v+\gamma & \text{if }v<-\gamma%
\end{array}
\right.  ,  \label{slam}
\end{equation}%
and for a  vector $\mathbf{v=(}a_{1},...a_{p})^{\prime },$ $\mathbf{S}%
_{\gamma }(\mathbf{v)}$ is the vector whose $i$ component is $S_{\gamma}(a_{i}).$}

{\normalsize The algorithm starts with initial values $(\boldsymbol{\beta }%
^{(0)},\boldsymbol{\Gamma }^{(0)}),$ with $||\boldsymbol{\beta }^{0}||_{2}=1$
and their iterative steps are as follows. Suppose that the  values of the parameters at step $j$%
,  $(\boldsymbol{\beta }^{(j)},\boldsymbol{\Gamma }^{(j)})$, have  already been computed
with $||\boldsymbol{\beta }^{(j)}||_2=1$. Then, $\boldsymbol{\beta }^{(j+1)}$
and $\boldsymbol{\Gamma }^{(j+1)}$ are defined by%
\begin{equation*}
	\widetilde{\boldsymbol{\beta }}^{(j+1)}=\mathbf{S}_{\lambda \eta _{j}}(%
	\boldsymbol{\beta }^{(j)}-\eta _{j}g_{i}(\boldsymbol{\beta }^{(j)})),\text{ }%
	\boldsymbol{\beta }^{(j+1)}=\frac{\widetilde{\boldsymbol{\beta }}^{(j+1)}}{||%
		\widetilde{\boldsymbol{\beta }}^{(j+1)}||_2},
\end{equation*}%
where $\eta _{j}$ is the value of $\eta$ minimizing $RG_{i}^{\ast }({\boldsymbol{\beta }}^{(j+1)})$. By (\ref{slam}) we have that each component of $\boldsymbol{\beta }^{(j+1)})$ is given by  

\begin{equation}
{{\beta }}^{(j+1)}_{i}=\left\{
	\begin{array}
		[c]{ccc}%
		{{\beta }}^{(j)}_{i}-\eta g_{i}^{\ast}(\boldsymbol{{\beta }}^{(j)}_{i})\ \  & \text{if}
		& |{{\beta }}^{(j)}_{i}-\eta g_{i}(\boldsymbol{\beta}^{(j)})|>\eta\lambda\\
		0 & \text{if} & \left\vert \beta_{i}^{(j)}-\eta g_{i}(\boldsymbol{\beta}%
		^{(j)})\right\vert \leq\eta\lambda.
	\end{array}
	\right.  .\ \label{m2a}%
\end{equation}
Moreover, according to equation
(\ref{gama2})%
\begin{equation*}
\boldsymbol{\Gamma }^{(j+1)}=\mathcal{G}(\boldsymbol{\beta }%
^{(j+1)}).
\end{equation*}
The procedure stops when $|RG_{i}^{\ast }\boldsymbol{(\boldsymbol{\beta }}%
^{(j+1)}\boldsymbol{\boldsymbol{)-}}RG_{i}^{\ast }\boldsymbol{(\boldsymbol{%
\beta }}^{(j})\boldsymbol{\boldsymbol{)|/|}}RG_{i}^{\ast }\boldsymbol{(%
\boldsymbol{\beta }}^{(j}\boldsymbol{\boldsymbol{)|\leq \varepsilon }}$,
where $\varepsilon $ is chosen according the desired accuracy.  }

{\normalsize When the loss function is $G_{1}$, the initial value $%
\boldsymbol{\beta }^{(0)}$ may be chosen as the vector corresponding to the
first redundancy analysis component. That is,  
as the largest eigenvalue of the matrix $\widehat{\boldsymbol{\Sigma}}_{%
\mathbf{x}_{t}\mathbf{x}_{t}}^{-1}\widehat{\boldsymbol{\Sigma }}_{\mathbf{x}%
_{t}\mathbf{y}_{t+h}}\widehat{\boldsymbol{\Sigma }}_{\mathbf{y}_{t+h}\mathbf{%
x}_{t}}$, where for any pair of multiple time series $(\mathbf{x}_{t},%
\mathbf{y}_{t})$, $1\leq t \leq T$, $\widehat{ \boldsymbol{\Sigma}}_{\mathbf{%
x}_{t}\mathbf{y}_{t+h}}\ $is the sample cross covariance matrix \ between $%
\mathbf{x}_{t}$ and $\mathbf{y}_{t+h}.$  }

{\normalsize When the loss function is $G_{2},$ an initial value for $%
\boldsymbol{\beta }^{(0)}$ may be \ taken \ as the vector corresponding to
the first canonical correlation. That is, as the largest eigenvector of
the matrix $\widehat{\boldsymbol{\Sigma }}_{\mathbf{y}_{t+h}\mathbf{y}%
_{t+h}}^{-1}\widehat{\boldsymbol{\Sigma }}_{\mathbf{y}_{t+h}\mathbf{x}_{t}}%
\widehat{\boldsymbol{\Sigma }}_{\mathbf{x}_{t}\mathbf{x}_{t}}^{-1}\widehat{%
\boldsymbol{\Sigma }}_{\mathbf{x}_{t}\mathbf{y}_{t+h}}.$  }

{\normalsize We have developed a computer program in R language to compute \
estimators of $\boldsymbol{\beta }$ and $\boldsymbol{\Gamma }$
using the loss function $G_{1}$. This program is used in the examples of
Section \ref{simulation}.  The derivative of $G_{1}^{\ast },$ required in (%
\ref{grad}) was computed numerically using the function grad in the R
package numDeriv. }

The following Theorem shows that in each step of the
algorithm described in this section the objective function RG$_{j}^{\ast}$,
$\ j=1,2$, decreases except when it reaches a critical point. Since the proof is
the same for both values of $j,$ this subscript is eliminated in the proof.   
  Let $I=\{1,2,...,p\}$ and  the vectors $\boldsymbol{\beta}$, $\mathbf{g}(\boldsymbol{\beta )}$, and $\mathbf{g}^{\ast }(\boldsymbol{\beta })$ 
  with components $\beta_{i}$,  ${g_{i}}(\boldsymbol{\beta })$ and ${g_{i}^{\ast }}(\boldsymbol{\beta })$ for $i=1,...,p$. Define  $I_{1}=\{i\in I:\beta_{i}=0$ and $|g_{i}(\boldsymbol{\beta}%
)|\leq\lambda\}$ and $I_{2}=\{i\in I:\beta_{i}^{{}}\neq0$ and $g_{i}^{\ast}(\boldsymbol{\beta})=0\}.$  Note that if  at the $jth$ step in the algorithm $\boldsymbol{\beta}^{(j)}$ has the $ith$ component   $\beta_{i}^{(j)}$ that belongs to either $I_{1}$ or $I_{2}$, the next value computed by the algorithm with  (\ref{m2a}) for this component,  $\beta_{i}^{(j+1)}$, will be zero. Therefore, if all the components of $\boldsymbol{\beta}^{(j)}$ belong to these two groups, that is, 
$\#I_{1}+\#I_{2}=p$, the algorithm will stop and we will say that  $\boldsymbol{\beta}^{(j)}$  is a critical point.

\begin{theorem}
Let $\boldsymbol{\beta}^{(0)}\boldsymbol{=(}\beta_{1}^{(0)},\beta_{2}%
^{(0)}..,\beta_{p}^{(0)})^{\prime}\in R^{p}$ be a non-critical point. Given
$\eta>0,$ let $\boldsymbol{\beta}^{(1,\eta)}=(\beta_{1,}^{(1,\eta)}%
,...,\beta_{p}^{(1,\eta)})^{\prime}=\mathbf{S}_{\eta\lambda}(\boldsymbol{\beta
}^{(0)}-\eta\mathbf{g}(\boldsymbol{\beta}^{(0)})).$ Then, there exists
$\eta_{0},$ such that for $\eta\leq\eta_{0}$ we have RG$^{\ast}%
(\boldsymbol{\beta}^{(1,\eta)})<$RG$^{\ast}(\boldsymbol{\beta}^{(0)}).$
\end{theorem}
{\normalsize \textbf{Proof} See the Appendix.}

 Let $\boldsymbol{\beta}^{(j)},1\leq j<\infty$ be the value obtained in the
$j-$th \ step of the algorithm described in this section. The following theorem shows that if  $\boldsymbol{\beta}_{0}$  is a limit point of this sequence, that is, if  there exists a subsequence
$\boldsymbol{\beta}^{(j_{k})},$ $1\leq k<\infty$ such that $\lim
_{k\rightarrow\infty}\boldsymbol{\beta}^{(j_{k})}=\boldsymbol{\beta}_{0}.$ Then
\ $\boldsymbol{\beta}_{0}$ is a critical point.    
\begin{theorem}
Suppose that the sequence  $\boldsymbol{\beta}^{(j)},1\leq j<\infty$ never reaches a critical point. Let $\boldsymbol{\beta}_{0}$ be a limit point
of this sequence, then it is a critical point.
\end{theorem}
{\normalsize \textbf{Proof} See the Appendix.}
 
\subsection{\protect\normalsize Non-regularized core component estimators.}
The next two theorems give the equations satisfied by the
estimators of $\boldsymbol{\beta}$ and $\boldsymbol{\Gamma}$ in the case
of non-regularized core components, that is when $\lambda=0$. It will be
shown how these equations can be used to derive a simple computational
sequential algorithm to compute these estimators. 

\begin{theorem}
{\normalsize When the loss function is given by (\ref{G1}) the optimal 
values $\widehat{\boldsymbol{\beta }}$ and $\widehat{\mathbf{\Gamma }}
$ satisfy the following fixed point equations
\begin{equation}
 \widehat{\boldsymbol{\beta }}=\left( \frac{1}{T-c-k-h}\sum_{t=c+k+1}^{T-h}%
\mathcal{X}_{t}^{\prime \ }\widehat{\mathbf{\Gamma }}^{\prime }\widehat{%
\mathbf{\Gamma }}\mathcal{X}_{t}\right) ^{-1}\left( \frac{1}{T-c-k-h}%
\sum_{t=c+k+1}^{T-h}\mathcal{X}_{t}^{\prime }\boldsymbol{\ }\widehat{\mathbf{%
\Gamma }}^{\prime }\mathbf{y}_{t+h}\right) ,  \label{eq1}
\end{equation}
and  
\begin{equation}
\widehat{\mathbf{\Gamma }}  =\mathcal{G}(
\widehat{\boldsymbol{\beta}}),  \label{eq20}
 \end{equation}
 where $ \mathcal{G}(\boldsymbol{\beta})$ is given in (\ref{Gammaa}).} 
\end{theorem}

{\normalsize \textbf{Proof} See the Appendix.}

\begin{theorem}The estimators $\widehat{\boldsymbol{\beta}} $ and $\widehat{ 
\mathbf{\Gamma}}$  minimizing the     loss function (\ref{G2}) satisfy the following fixed point  equations
{\normalsize Let 
\begin{equation}
\widehat{\mathbf{\Sigma }}_{{}}=\frac{1}{T-c-k-h}\sum_{t=c+k+1}^{T-h}\left[ (%
\mathbf{y}_{t+h}-\widehat{\mathbf{\Gamma }}\mathcal{X}_{t}\widehat{%
\boldsymbol{\beta }})(\mathbf{y}_{t+h}-\widehat{\mathbf{\Gamma }}%
\mathcal{X}_{t}\widehat{\boldsymbol{\beta }})^{\prime }\right].
\label{eq5}
\end{equation}%
}	
Then,  
{\normalsize 
\begin{equation}
\widehat{\boldsymbol{\beta }}=\left[ \frac{1}{T-c-k-h}\sum_{t=k+c+1}^{T-h}(%
\mathcal{X}_{t}^{\prime }\boldsymbol{\ }\widehat{\mathbf{\Gamma }}%
^{\prime }\widehat{\mathbf{\Sigma }}^{-1}\widehat{\mathbf{\Gamma }}%
\mathcal{X}_{t})^{-1}\right] \left[ \frac{1}{T-c-k-h}\sum_{t=c+k+1}^{T-h}(%
\mathcal{X}_{t}^{\prime }\boldsymbol{\ }\widehat{\mathbf{\Gamma }}%
^{\prime }\widehat{\mathbf{\Sigma }}^{-1}\mathbf{y}_{t+h})\right] 
\label{eq3}
\end{equation}%
and}%
\begin{equation}
\widehat{\mathbf{\Gamma }}=\mathcal{G}(\widehat{\boldsymbol{\beta }}^{\ })\label{TE22}.
\end{equation}
\end{theorem}
{\normalsize \textbf{Proof} See the Appendix.}

A simple sequential algorithm for the non-regularized estimators based on core components when the loss function is $G_1$
can be summarized as follows. In step 1, an initial value  $\boldsymbol{\beta^{(1)}}$  is given.
Suppose that   $i$ steps of the algorithm  have been  executed, and in this step the value of the parameter $\boldsymbol{\beta}$ is  $\boldsymbol{\beta^{(i)}}$. Then,    $\boldsymbol{\beta^{(i+1)}}$ is obtained as the left hand right side of (\ref{eq1}) with $\widehat{\boldsymbol{\beta}}$  replaced by $\boldsymbol{\beta}^{(i)}$ and $\widehat{\boldsymbol{\Gamma}}$ by   $\mathcal{G}(\boldsymbol{\beta^{(i)}})$. The procedure stops at the first value $i$ such that $||  \boldsymbol{\beta}^{i+1}-\boldsymbol{\beta}^{i}||_2\leq \varepsilon$. Call to this value $i_0$, then $\widehat{\boldsymbol{\beta}}=\boldsymbol{\beta}^{(i_0+1)}$ and $\widehat{\boldsymbol{\Gamma}}=\mathcal{G}(\widehat{\boldsymbol{\beta}})$. The algorithm for the case that $G_2$ is minimized  is completely similar.

\section{\protect\normalsize Asymptotics}  \label{asymp}

{\normalsize In this Section we state a theorem that asserts under some conditions the almost
surely convergence of   the  non-regularized estimators ($\lambda=0$) of the  core components, which are the values of $\boldsymbol{\beta}$ and $\boldsymbol{\Gamma}$ that minimize the function $G_1$ given in (\ref{G1}),  to their population values.  Let $\widehat{%
\boldsymbol{\beta }}_{T}^{(j)}$ and $\widehat{\Gamma }_{T}^{(j)}$ be the values of these estimators 
   corresponding to the $j$-th \ core\ component using lags $%
c_{j}$ and $k_{j},$ $1\leq j\leq s$, and sample size $T$. We assume that the lag parameters $c_j$, $k_j$ and the number  of components $s$ will remain fix for all sample size $T$. Let $\mathbf{v}%
_{t}=(\mathbf{y}_{t}^{\prime },\mathbf{z}_{t}^{^{\prime }})^{\prime }$, $%
t\in \mathbb{Z},$ where $\mathbb{Z}\mathbf{\ }$is the set of integer
numbers. For $1\leq j\leq s$, put $\mathbf{x}_{t}^{(j)}=(\mathbf{z}%
_{t}^{\prime },\mathbf{z}_{t-1}^{\prime },...,\mathbf{z}_{t-c_{j}}^{\prime
})^{\prime }$. Note that \ $\mathbf{x}_{t}^{(j)}\in R^{p_{j}},$ where $%
p_{j}=m(c_{j}+1)$.  }

{\normalsize We will assume the following properties.  }

{\normalsize P$1$. The series $\mathbf{v}_{t}$ is a strictly stationary \
and ergodic. Moreover $E(\mathbf{v}_{t})=\mathbf{0}$ and $E(||\mathbf{v}%
_{t}||_{2}^{2})<\infty .$  }

{\normalsize The property P2 defines the population values of the estimators
required for the $s $ core components and it is stated recursively for $%
1\leq j\leq s.$  }

{\normalsize P$2$. For the first component, given $\boldsymbol{\beta }\in
R^{p_{_{1}}}$ with $||\boldsymbol{\beta ||}_{2}=1$ and $\boldsymbol{\Gamma }$
a $q\times (k_{1}+1)$ matrix, put $f_{t}^{(1)}(\boldsymbol{\beta })=%
\boldsymbol{\beta }^{\prime }\mathbf{x}_{t}^{(1)},$ $\mathbf{f}_{t}^{(1)}(%
\boldsymbol{\beta }$ $)=(f_{t}^{(1)}(\boldsymbol{\beta )}$,$f_{t-1}^{(1)}(%
\boldsymbol{\beta )},...,f_{t-k_{1}}^{(1)}(\boldsymbol{\beta )})^{\prime }$.
Then, given the parameters $\boldsymbol{\beta }$ and $\boldsymbol{\Gamma }$,
the expected squared error for the first component \ is ESE$^{(1)}(%
\boldsymbol{\beta },\boldsymbol{\Gamma })=E(||(\mathbf{y}_{t+h}\mathbf{-}%
\boldsymbol{\Gamma }\mathbf{f}_{t}^{(1)}\mathbf{(}\boldsymbol{%
\beta }))||_{2}^2)$. We assume that there exists a unique $(\boldsymbol{%
\beta}$, $\boldsymbol{\Gamma})$ that minimizes ESE$^{(1)}(\boldsymbol{\beta},%
\boldsymbol{\ \Gamma })$}.

{\normalsize Then, the population values of the estimators $\widehat{%
\boldsymbol{\beta }}_{T}^{(1)}$ and $\widehat{\boldsymbol{\Gamma }}%
_{T}^{(1)}$ are 
\begin{equation*}
(\boldsymbol{\beta }_{\ast }^{(1)},\boldsymbol{\Gamma }_{\ast
}^{(1)})=\arg _{(||\boldsymbol{\beta ||}_{2}=1,\boldsymbol{\Gamma })}\min 
\text{ESE}^{(1)}(\boldsymbol{\beta },\boldsymbol{\Gamma })
\end{equation*}%
and the first population residual error is $\mathbf{e}_{\ast t}^{(1)}=\mathbf{y}_{t}-%
\boldsymbol{\Gamma }_{\ast }^{(1)}\mathbf{f}_{t-h}^{(1)}\mathbf{(}%
\boldsymbol{\beta }_{\ast }^{(1)}).$  }

{\normalsize Suppose that we have already defined the residual errors $%
\mathbf{e}_{\ast t}^{(j)}\mathbf{\ }$with $j<s.$ Then, given $\boldsymbol{\beta }$
$\in R^{p_{j+1}}$ and a $q\times (k_{j+1}+1)$ matrix $\boldsymbol{\Gamma ,}$
put $\ f_{t}^{(j+1)}(\boldsymbol{\beta }$ $)=\boldsymbol{\beta }$ $^{\prime }%
\mathbf{x}_{t}^{(j+1)},$ $\mathbf{f}_{t}^{(j+1)}(\boldsymbol{\beta }%
)=(f_{t}^{(j+1)}$,$f_{t-1}^{(j+1)},...,f_{t-k_{j}}^{(j+1)})^{\prime }$ and
ESE$^{(j+1)}(\boldsymbol{\beta }$ $,\boldsymbol{\Gamma })=E(||(\mathbf{e}%
_{\ast t+h}^{(j)}\mathbf{\ -}$\textbf{$\boldsymbol{\Gamma }$}$\mathbf{f}_{t}^{j+1}%
\mathbf{(}\boldsymbol{\ \beta }$ $))||_{2}^2)$. {It is assumed that there
exists a unique $(\boldsymbol{\beta }$, $\boldsymbol{\Gamma })$ that
minimize ESE$^{(j+1)}(\boldsymbol{\beta },\boldsymbol{\Gamma })$ }. Then,
the population values for the estimators $\widehat{\boldsymbol{\beta }}%
_{T}^{(j+1)}$ and $\widehat{\boldsymbol{\Gamma }}_{T}^{(j+1)}$ are %
\begin{equation*}
(\boldsymbol{\beta }_{\ast }^{(j+1)},\boldsymbol{\Gamma }_{\ast
}^{(j+1)})=\arg _{(||\boldsymbol{\beta ||}_{2}=1,\boldsymbol{\Gamma })}\min 
\text{ESE}^{(j+1)}(\boldsymbol{\beta },\boldsymbol{\Gamma })
\end{equation*}%
and the $(j+1)$-th residual error is $\mathbf{e }_{\ast t}^{(j+1)}=\mathbf{e}%
_{\ast t}^{(j)}\mathbf{-}$\textbf{$\boldsymbol{\Gamma }$}$_{\ast }^{(j+1)}%
\mathbf{f}_{t-h}^{(j+1)}\mathbf{(}\boldsymbol{\beta }_{\ast }^{(j+1)}).$  }

\begin{theorem}
{\normalsize Assume that P$1$, and P$2\ $hold. Then \ $\lim_{T\rightarrow
\infty }$ $\widehat{\boldsymbol{\beta }}_{T}^{(j)}=\boldsymbol{\beta }%
_{\ast }^{(j)}$ and $\lim_{T\rightarrow \infty }\widehat{\boldsymbol{\Gamma 
}}_{T}^{(j)}=\boldsymbol{\Gamma }_{\ast }^{(j)}$ almost surely for $1\leq
j\leq r.$  }
\end{theorem}

{\normalsize The proof of this theorem is similar \ to Theorem 2 \ of \cite%
{pena2019}. The only difference is that the vector to predict is $\mathbf{y}%
_{t}$ instead of $\mathbf{z}_{t}$ and that in order to define sequentially \
the population values of the parameters, it is assumed the uniqueness of the
population parameters $(\boldsymbol{\beta }_{\ast }^{(j)},\boldsymbol{%
\Gamma }_{\ast }^{(j)}),$ $1\leq i\leq r.$ In a similar way\ it \ can be
defined the population values of the estimator when they are defined by (\ref%
{mRG2}) and to prove consistency.}

\section{\protect\normalsize The Core Component Forecasting (CCF) Procedure
with Cross-Validation}

{\normalsize \label{cross} }

{\normalsize In this section we describe the Core Components Forecasting 
(CCF) procedure using the regularized estimators (RCC) defined in Section  %
\ref{CCE}. Given $(\mathbf{y}_{t}^{\prime },\mathbf{z}_{t}^{\prime
})^{\prime }$, $1\leq t\leq T$, with $\mathbf{y}_{t}\in R^{q}$, and $\mathbf{%
\ z}_{t}\in R^{m}$, {\ in order to forecast $\ \mathbf{y}_{T+h}$ we need to
compute the RCC-estimators $\widehat{\boldsymbol{\beta}} $ and $\widehat{%
\mathbf{\Gamma}}^{{}}$, given the values of the hyper-parameters: the
number of lags $c_{i}$ and $k_{i}$ and the  penalization parameter $\lambda_{i} $ for
each component, $1 \leq i \leq s$, and the number of core components to be used, $s$. By the
sake of simplicity we will only consider  the loss function $G_{1}$,
although the case of $G_{2}$ is completely similar.} }

{\normalsize The forecast of $y_{T+h}$ \ is obtained  throughout \ a
sequential procedure. We divide the complete sample in one  training part
and two validation parts. Let $0<\alpha <1$ and $T_{1}=$ $[\alpha T],$ then
the training data includes the observations $ (\mathbf{y}_{t}^{\prime },%
\mathbf{z}_{t}^{\prime })^{\prime }$, $1\leq t\leq  T_{1}$, where $[u]$
denotes the integer part of $T_{1}$. {\ The training data will be used to
compute the RCC-estimators $\widehat{\boldsymbol{\beta}}^{(1)} $ and $%
\widehat{\mathbf{\Gamma}}^{(1)}$, given the values of the
hyper-parameters. The  \ first validation set is composed for the following $%
T_{2} =[(1-\alpha )T/2]$ observations  and is used to determine the
hyper-parameters ($c_{i}$, $k_{i}$, $\lambda_{i}$) for each component, and for selecting
the number of components $s$. The second validation set is composed by the
remaining $T_{3}=(T-T_{1}-T_{2})\ $ observations and will be used to
obtain an unbiased estimator of the \ forecasting mean squared error (FMSE)
of the whole procedure. \ The maximum values of the hyper-parameters $c$,$k$%
, $\lambda $ for all  the core components are fixed as \ $c_{\max },k_{\max }
$ and $\lambda _{\max }  $. \ In the following numbered list we described
the steps to compute the  CCF procedure .} }

\begin{enumerate}
 \item {\normalsize We start with the first component. Let $\widehat{\boldsymbol{%
\beta }}^{(1)} _{c,k,\lambda },\widehat{\boldsymbol{\Gamma }}%
_{c,k,\lambda }^{(1)}$ be the RCC-estimators using the training set \ with
lag  parameters $c$ and $k$ and penalization parameter $\lambda $ and
forecasting horizon $h$. Let  
 \begin{equation*}
\text{FMSE}_{c,k,\lambda }=\frac{1}{(T-h-c-k)}
\sum_{t=T_{1}+1}^{T_{1}-T_{2}}(||\mathbf{e}_{t+h|t}( \widehat{\boldsymbol{%
\beta }} _{c,k,\lambda }^{(1)},\widehat{\boldsymbol{\Gamma }}%
_{c,k,\lambda }^{(1)}|| _{2}^{2})
\end{equation*}
\ a cross validated estimator of the FMSE. The  optimal values of $c$ and $k$ that
will be used to estimate the first core component  are given by 
\begin{equation}
(\widehat{c}_{1},\widehat{k}_{1})=\arg \min_{0\leq c\leq c_{\max },0\leq k\leq k_{\max }} 
\text{FMSE}_{c,k,0}.  \label{ckop}
\end{equation}}

 {\normalsize To choose the penalization parameter $\lambda $ for the first 
core component we will compute $(\widehat{\boldsymbol{\beta }}^{(1)}
_{\widehat{c}_{1},\widehat{k}_{1},\lambda } ,\widehat{\boldsymbol{\Gamma}}^{(1)}
_{\widehat{c}_{1},\widehat{k}_{1},\lambda }) $  in the  grid of $J$ elements $%
\lambda _{j}=j\lambda _{\max }/(J-1),0\leq j\leq (J-1).$  Then, a cross validated 
estimator of $\lambda_1$ is given by 
\begin{equation*}
\widehat{\lambda}_{1}=\arg \min_{0\leq j\leq J-1}\ \text{FMSE}
_{\widehat{c}_{1},\widehat{k}_{1},\lambda _{j}} 
\end{equation*}
and the estimators  $\boldsymbol{\beta}^{(1)   }$ and $\boldsymbol{\Gamma}^{(1)}$ for the first components are   $ 
 \widehat{\boldsymbol{\beta }}^{(1)}
_{\widehat{c}_{1},\widehat{k}_{1},\widehat{\lambda }_1}$ and  $ 
 \widehat{\boldsymbol{\Gamma }}^{(1)}
_{\widehat{c}_{1},\widehat{k}_{1},\widehat{\lambda }_1}$respectively. A cross validated FMSE estimator when using only one core  component
can be estimated by 
\begin{equation*}
\text{FMSE}^{(1)}=\ \text{FMSE}_{\widehat{c}_{1},\widehat{k}_{1},\widehat{\lambda}_{1}}.
\end{equation*}
}

\item {\normalsize Suppose that the \ optimal values $\widehat{c}_{v}$, $\widehat{k}_{v}$,
	 $\widehat{\lambda} _{v}$,  for $1\leq v\leq s$, have been computed, as well 
as the corresponding estimators of $\boldsymbol{\beta }$ and $\boldsymbol{\
\Gamma }$ using the training sample. Let $\widehat{\mathbf{e}}
_{t|t-h}^{(s)}$ be the vector of forecasting errors of $\mathbf{y}_{t}$ \ 
when \ it is predicted using these $s$ core components $.$ Then, the optimal
values for the $(s+1)$-th components \ $\widehat{c}_{s+1}$, $\widehat{k}_{s+1}$,
 $\widehat{\lambda}_{s+1}$, $\widehat{\boldsymbol{\beta }} ^{(s+1)},\widehat{%
\boldsymbol{\Gamma }}^{(s+1)}$ and $\ $\ the cross  validated
forecasting MSE \ using $s+1$ components FMSE$ ^{(s+1)},$ are obtained as
for $s=1,$ but replacing the $ \mathbf{y}_{t}$ s for $\widehat{\mathbf{e}}%
_{t|t-h}^{(s)}$.  }
 
\item {\normalsize We continue with this process of computing \ the optimal 
values $\widehat{c}_{s},$ $\widehat{k}_{s}$, $\widehat{\lambda}_{s}$, and FMSE$^{(s)}$  until for
the first time $\ $ FMSE$^{(s+1)}\geq $FMSE$^{(s)}$ . Then the number of
components is chosen equal to $s$. This value will be denoted by $\widehat{s}$. }

\item 
{\normalsize Note that as FMSE$^{(s)}$ is based on the observations $( 
\mathbf{y}_{t}^{\prime },\mathbf{z}_{t}^{\prime })^{\prime },$ $T_{1}+1\leq 
t\leq T_{1}+T_{2}$, and these observations were used to determine the hyper-parameters
$\widehat{c}_{s},\widehat{k}_{s}$, and $\widehat{\lambda}_{s}$. Therefore, it \  \ is not an unbiased
estimator of the FMSE of the described  procedure. An unbiased estimator \
may be obtained by cross  validation using $(\mathbf{y}_{t}^{\prime },$ $%
\mathbf{z} _{t}^{\prime })^{\prime },T_{1}+T_{2}+1\leq t\leq T,$ that is  by 
\begin{equation*}
\text{FMSECV}=\frac{1}{(T-h-c-k)}\sum_{t=T_{1}+T_{2}+1}^{T}||\mathbf{e}%
_{t+h}^{\left( \widehat{s}\right) }||_{2}^{2}.
\end{equation*}
This is the value that will be used to make comparisons with other 
prediction procedures. }
 
\item {\normalsize The final forecast of $\mathbf{y}_{T+h}$ is made by using
(\ref{predfor}) with $\widehat{s}$ components and where for the $i$ 
component $(\boldsymbol{\beta } \boldsymbol{,\Gamma }  
\boldsymbol{)}$  is estimated  using  the  values $%
c=\widehat{c}_{i}, \ k=\widehat{k}_{i} $ and $\lambda =\widehat{\lambda}_{i}.\ $  }
\end{enumerate}
 
 \section{\protect\normalsize Problems Covered by the General Dimension
Reduction Approach}

{\normalsize \label{PCCCFA} }

\subsection{\protect\normalsize One-sided dynamic principal components for
forecasting (DPCF)} {\normalsize \label{DPCF} 
Dynamic principal components (DPC) were introduced
	by \cite{brillinger64}, as two sided linear combinations of the series and
	their lags that provide an optimal reconstruction of the observed set of
	series by using the present values and all their lags. He showed how these
	components can be computed in an infinite realization of a stationary process
	using frequency domain tools. The properties of these components have been
	used to estimate the factors and generate forecasts in generalized dynamic
	factor models, see \cite{forni2005}.  }

{\normalsize   In the one sided dynamic principal components (OSDPC) approach
	proposed in \cite{pena2019} the available data are a vector time series
	$\mathbf{y}_{t}=(y_{t,1},...,y_{t,q}),$ $1\leq t\leq T,$ to be predicted with
	some horizon $h$. This can be written as a particular case of the structure
	assumed in this article for the core components by putting $\mathbf{z}%
	_{t}=\mathbf{y}_{t}$ and $\mathbf{x}_{t}=(\mathbf{y}_{t},\mathbf{y}%
	_{t-1},...\mathbf{y}_{t-c_{i}})$ , where $c_{i}$ is the number of lags of
	$\mathbf{y}_{t}$ to define the $i$th principal component. We also have to
	determine $k_{i}$, the number of lags of the $i$-th dynamic principal component
	that will be used to forecast $\mathbf{y}_{t+h}$ in (\ref{pred1}). The main difference 
	between both approaches is that 
	in the OSDPC approach the parameters that define the linear combination used
to construct the $i$-th component, $\boldsymbol{\beta}^{(i)},$  and the
loadings of the $i$-th component in the forecast, $\boldsymbol{\Gamma}^{(i)},$
are  assumed to be the same for any forecasting horizon $h$ and the forecast are {\normalsize
	generated by estimating ARIMA\ models for the components and using these
	forecasting values in the prediction. Therefore, it is assumed an underline
	linear model for the vector time series.  However, in the core procedure a linear model 
	is not assumed and, therefore, a
	different model is fitted for each forecasting horizon allowing for some non
	linear change in the future evolution of the vector series. }}

 \subsection{\protect\normalsize Dynamic Canonical Correlation (DCC)}
{\normalsize Canonical correlation was proposed first by independent data
but has found  many applications in time series. See, for instance, \cite%
{robinson1973}, \cite{akaike1976canonical},  \cite{tsay1985use}, or \cite%
{johansen1988}.  Suppose that we have a sample of two random vectors $(%
\mathbf{x}_{i},\mathbf{y}_{i}),$ $1\leq i\leq N.$ The first canonical
correlations variables are defined as two scalar variables, $\boldsymbol{%
\beta }^{(1)\prime }\mathbf{x}_{i}$ \ and $\boldsymbol{\gamma }^{(1)\prime }%
\mathbf{y}_{i},$ with maximum square correlation coefficient and with the
constraint that \ $||\boldsymbol{\beta }||_{2}=1,||\boldsymbol{\gamma }%
||_{2}=1$, that is, 
\begin{equation*}
(\boldsymbol{\beta }^{%
{\acute{}}%
(1)},\boldsymbol{\gamma }^{(1)})=\arg \sup_{||\boldsymbol{\beta }||_{2}=1,||%
\boldsymbol{\gamma }||_{2}=1}\rho ^{2}(\boldsymbol{\beta }^{\prime }\mathbf{x%
}_{i},\boldsymbol{\gamma }^{\prime }\mathbf{y}_{i}),\text{ }
\end{equation*}
where $\rho $ is the empirical correlation coefficient.  }

{\normalsize Higher order canonical correlations are defined recursively.
Suppose that the $j$-th canonical correlation variables, $(\boldsymbol{\beta 
}^{(j)\prime }\mathbf{x}_{i},\boldsymbol{\gamma }^{(j)\prime }\mathbf{y}%
_{i}) $ are already defined for $1\leq j\leq s.$ Then, the $(s+1)$-th two
canonical correlation variables are defined by 
\begin{equation*}
(\boldsymbol{\beta }^{(r+1)},\boldsymbol{\gamma }^{(r+1)})=\arg \sup 
_{\substack{ ||\boldsymbol{\beta }||_{2}=1,||\boldsymbol{\gamma }||_{2}=1 
\\ \boldsymbol{\beta }\perp \boldsymbol{\beta }^{(i)},\boldsymbol{\gamma }%
\perp \boldsymbol{\gamma }^{(i)},1\leq i\leq r,}}\rho ^{2}(\boldsymbol{\beta 
}^{\prime }\mathbf{x}_{i},\boldsymbol{\gamma }^{\prime }\mathbf{y}_{i}).%
\text{ }
\end{equation*}%
Let $\ $ $s\leq \min (p,q).$ An interesting property of the canonical
correlations variables is that \ for $1\leq r\leq \min (p,q)$ they coincide
with some corresponding static core components with all the values of $c$, $k
$ ,$\lambda$ and $h$ equal to $0$ \ but and computed with the loss function $%
G_{2}$. (see \cite{yohai1980canonical}). A similar property holds if we
reverse the roles of $\mathbf{y}$ and $\mathbf{x}$. }

{\normalsize Based on this result, we propose a forecasting oriented
definition of dynamic canonical correlation time series variables by
generalizing this prediction property of the static canonical correlation.
Given two finite vector time series $\mathbf{z}_{t}$ and $\mathbf{w}_{t},$ $%
1\leq t\leq T,$ \ with dimensions $p$ and $\ q$ and any $s\leq \min (p,q),$
we define the $s$-th canonical variable corresponding to $\mathbf{z}_{t}$ \
as follows. \ We apply the CCF procedure defined in Section \ref{cross} \
with loss $G_{2}$ up to the $s$ \ core component to predict $\mathbf{w}_{t+h}
$ using $\mathbf{z}_{t}$. Let $\mathbf{x}_{t}=(\mathbf{z}_{t},\mathbf{z}%
_{t-1},...,\mathbf{z}_{t-c_{\text{opt}}^{(r)}}),$ then the $s$-th dynamic
canonical correlation is $\ a_{t,h}=$ $\widehat{\boldsymbol{\beta }}%
^{(r)\prime }\mathbf{x}_{t}.$\ The dynamic canonical correlations to
predict $\mathbf{z}_{t}$ based on $\mathbf{w}_{t}$ are similarly defined. }
 
\subsection{Dynamic Redundancy Analysis (DRA)}
Redundancy Analysis (RA) has been proposed as an alternative to canonical 
correlation analysis, see \cite{van1977redundancy}, by considering the two
set of variables symmetrically.  However, unlike canonical analysis, it has
not been often applied for time series. In this case we have two vector
series, $\mathbf{y}_{t}$ and $\mathbf{z}_{t}$, $ 1\leq t\leq T$,
of dimensions $q$ and $m$ respectively. The goal is to find few linear
combinations of $\mathbf{x}_{t}=(\mathbf{z}_{t}$ $\mathbf{z}_{t-1}....,,%
\mathbf{z}_{t-c}),$ where $p=m(c+1)$, to predict 
\begin{equation*}
\mathbf{y}_{t+h}=({y}_{t+h,1}...,{y}_{t+h,q}),
\end{equation*}%
so that the FMSE is minimized. We may also include in the set $\mathbf{x}_{t}
$ lag values of the vector of variables to be forecast. This is a particular
case of our general set up and we may define the dynamic redundancy analysis
solution as given by equation (\ref{eq1}) and (\ref{eq20}).
 
\section{{\protect\normalsize A Monte Carlo Example\label{simulation}}}

{\normalsize We perform a\ Monte Carlo simulation to compare the proposed
CCF procedure defined in Section \ref{cross} , based on loss function $G_1$,
to the forecasting procedure of \cite{gao2024supervised} described in
Section \ref{intro}, that will be indicated as the sdPCA procedure. They are
compared generating the multiple series $(\mathbf{z}%
_{t,},\mathbf{y}_{t}),1\leq t$ $\leq 200+1$, both of dimension  50, and forecast horizon $h=1$, so that the \ series $\mathbf{x}%
_{t}=(\mathbf{z}_{t},\mathbf{\ z}_{t-1},...,\mathbf{z}_{t-k})$ will be used
to predict $\mathbf{y}_{t+1}$. Also, before applying the forecasting
procedures, all the components of $\mathbf{z}_{t}$ and $\mathbf{y}_{t}$ are
standardized to mean zero and variance one. The number of lags, $k$, and $c$ the
value of $\lambda$ for each component and the optimal number of components, $%
s$, will be chosen using cross validation as described in Section \ref{cross}, and 
 the number of lags and components for the sdPCA procedure were similarly determined.
The training set where the parameters are
initially estimated consists of the first 140 observations, $70\%$ of the
data. The first testing set, with the following 30 observations, is used to obtain
the values of $k$ ,$c$ and $s$ for the sdPCA procedure. Finally, the parameters of the procedures are
estimated with the first 200 observations and, using these parameters, $%
\mathbf{y}_{201}$ is predicted. 500 replications are generated.}

 Let $\mathbf{y%
}_{201}^{(i)}$ be the value of $\mathbf{y}_{201}$ in the $i$-th replication,
and $\widehat{\mathbf{y}}_{201}^{(i)}$ its prediction. The performance of
each procedure is measured by the forecasting mean squared error defined by  

{\normalsize 
\begin{equation*}
\text{FMSE}=\frac{1}{500}\sum_{i=1}^{500}\left\Vert \mathbf{y}_{201}^{(i)}-%
\widehat{ \mathbf{y}}_{201}^{(i)}\right\Vert ^{2}_2.
\end{equation*}
}

{\normalsize The vectors $\mathbf{z}_{t}$ and $\ \mathbf{y}%
_{t}\ ,1\leq t\leq201$ are generated as follows.
First, the explanatory variables, $\mathbf{z}_{t}$, are generated as a
vector time series that follows a VAR(1) model of the form 
\begin{equation*}
\mathbf{z}_{t}=\mathbf{Az}_{t-1}+\boldsymbol{\varepsilon}_{t},
\end{equation*}
where $\mathbf{A}$ is a $50\times50$ matrix such that $\mathbf{A}%
(i,i)=d_{i}, $ $1$ $\leq i\leq50,$ $\mathbf{A}(i,i+1)=g_{i}\mathbf{,}$ $%
1\leq i\leq49$ and its remaining elements are $0,$ and $\mathbf{d}%
=(d_1,...,d_{50})$, and $\mathbf{g}=(g_1,...,g_{49})$ are vectors with all
its components generated as independent uniform random variables in the
interval [0,0.5]. The $\boldsymbol{\varepsilon}_{t}$\'{}s \ are independent
random vectors with multivariate normal distribution \ with mean $\mathbf{0}$
and covariance matrix diag$(\mathbf{v})$, where $\mathbf{v}$ is a vector of
dimension 50 of independent uniform random variables in [0,0.5]. }

{\normalsize Then, a factor time series, $f_{t}$, is generated using these
explanatory variables as 
\begin{equation*}
f_{t}=\mathbf{b}_{1}^{\prime}\mathbf{z}_{t}+\mathbf{b}_{2}^{\prime}\mathbf{z}
_{t-1},
\end{equation*}
where $\mathbf{b}_{1}$ and $\mathbf{b}_{2}$ are independent N$(0,1)$ random
variables. }

{\normalsize The series to be forecasted, $\mathbf{y}_{t}$, follows a
generalized dynamic factor model with a single factor, 
\begin{equation*}
\mathbf{y}_{t}=\mathbf{c}_{1}f_{t}+\mathbf{c}_{2}f_{t-1}+\mathbf{e}_{t},
\end{equation*}
where the vectors $\mathbf{c}_{1},\mathbf{c}_{2}$ are of dimension $50,$
with all their elements independent N$(0,1)$ random variables and the $%
\mathbf{e}_{t}^{{}}$s are \ i.i.d. random vectors with multivariate normal
distribution with mean $\mathbf{0}$ and covariance matrix equal to $%
\sigma_{e}^2\mathbf{I}$. }

{\normalsize Table \ref{tab:ex1} presents the results of comparing in this
example the one-step-ahead FMSE of the CCF defined in Section \ref{cross}
and the sdPCA procedures as a function of the standard deviation of the noise
of the $y_t$ variables. The  smaller the value of $\sigma_{e}$ the larger
the signal versus the noise in the responses. It can be seen that, in this
example, the CCF procedure works better in all cases than the sdPCA procedure that has a larger FMSE in the range from 50\% and 34\%.  }

\begin{table}[h]
	\caption{MEFE of the sdPCA and CCF procedures in the Example}
	\label{tab:ex1}
	\centering	
	\begin{tabular}{c|ccccc}
		$\sigma_{e}$       & 0.3    & 0.7 &  1 & 2 & 3\\ \hline
		$\text{FMSE(sdPCA)}$ &  0.6691 &0.6705    & 0.6792 &  0.7039 &0.7332\\ 
		$\text{FMSE(CCF)}$ &  0.4457 &  0.4593  & 0.4662 &  0.5041 & 0.5472\\ 
		FMSE(sdPCA)/FMSE(CCF) & 1.501  & 1.460  & 1.456  & 1.396 & 1.340\\ \hline
	\end{tabular}
\end{table}

\section{\protect\normalsize \ A Forecasting Application}

{\normalsize We consider a set of time series of cost of living index in
Europe. The variables to be forecasted, that form the vector $\mathbf{y}_{t}$%
, are the monthly harmonized index of consumer prices of the 20 european
countries that have the euro as currency, the Eurozone countries. The set of
explanatory variables, $\mathbf{z}_{t}$, is composed for all the variables
of $\mathbf{y}_{t}$, the indexes for other seven european countries that are
not in the Eurozone, but belong, or have an association, with the European
Union (Bulgari, Chequia, Denmark, Island, Norway, Romania and Sweden) and
two global indexes built by Eurostat (the European Statistical Office), one
for the whole European Union (index EU) and other for the Euro Area (index
EA). Thus, in this example $q=20$ and $m=29$. Note that the countries
included in these global indexes have been changing during the period
considered: in the XXI century the Euro area has increased from 12 to 20
countries and the European Union from 25 to 27 members. The data can be
downloaded from the Eurostat database, at
https://ec.europa.eu/eurostat/web/hicp/database. We will use a data set from
1/2001 to 12/2022, $T=264$. A plot of the 20 series to be predicted are
shown in figure \ref{fig:ECPseries} and a plot of the nine explanatory time
series are shown in Figure \ref{fig:NECPseries}. 
\begin{figure}[h]
\centering
\includegraphics[width=18cm, height=8cm]{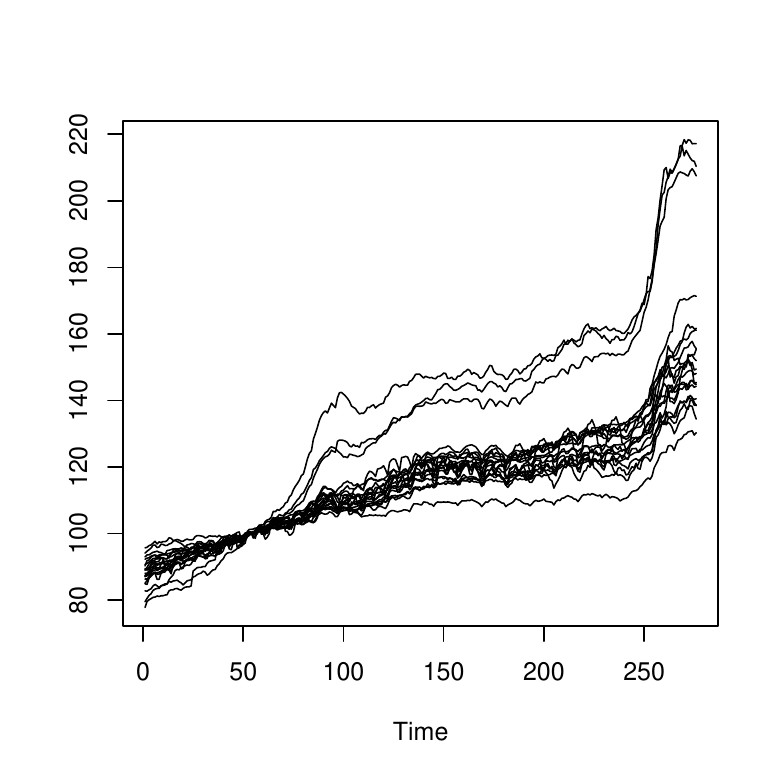} 
\caption{Harmonised index of consumer prices of the Eurozone countries
between 1/2001 to 12/2022}
\label{fig:ECPseries}
\end{figure}
}

\begin{figure}[h]
\centering
\includegraphics[width=18cm, height=8cm]{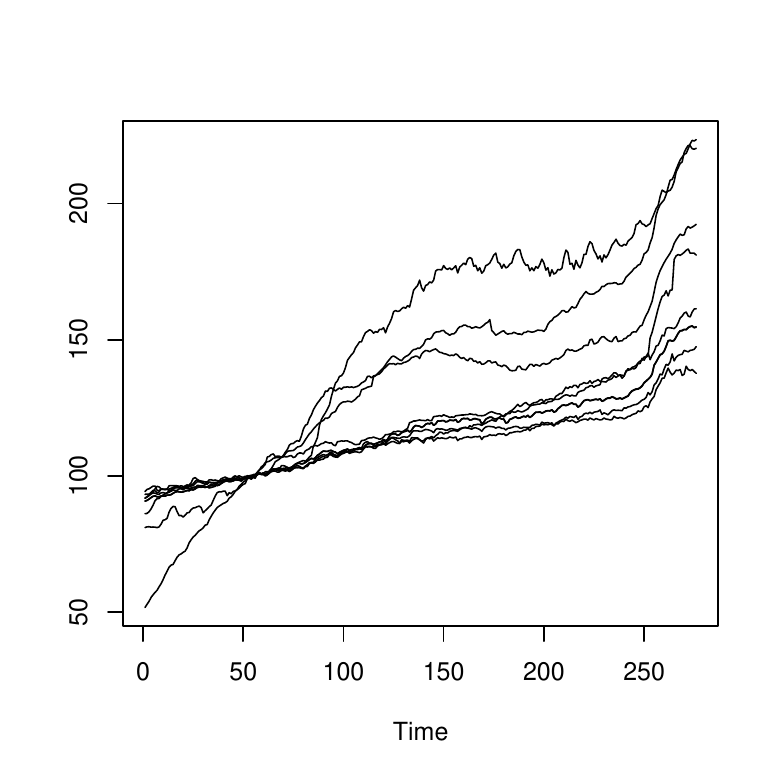}
\caption{Harmonised index of consumer prices of the Non Eurozone countries
between 1/2001 to 12/2022}
\label{fig:NECPseries}
\end{figure}

{\normalsize The forecasting performance of the sdPCA and CCF procedures have
been compared in this data as follows. For the CCF procedure  $\boldsymbol{\beta }$
and $\boldsymbol{\Gamma }$ for each component and for  different lag values $c$
and $k$ are estimated from the data using the first 70\% of the sample, 184
observations. The next 15\% observations, 40 observations in this case, are
used for estimating by cross validation the number of lags $k$ and $c$ used
for each components, and the number of components, $s$.  The final 15\% of observations, 40 data
in our case, are used for estimating using cross validation the FMSE of the
two procedures. The estimation of the FMSE of the sdPCA procedure was similarly obtained. 
The estimates of the FMSE for one step ahead forecasting  with this data are 0.0250 for the sdPCA and 0.0202 for the CCF}.

\section{\protect\normalsize Conclusions}

{\normalsize We have introduced a general formulation for forecasting a
vector time series using a set of high dimensional time series by using core
components. These core components are linear function of a large set of
explanatory variables, and are computed jointly with their coefficients in
the forecasting equation minimizing the prediction mean square error with a
lasso regularization. Our method takes a different approach from other
forecasting methods using factors, as those based on classical principal
components or scaled principal components, where, first, some components are
found with some criterion and, second, their coefficients in the forecasting
equation are obtained by minimizing the prediction error. However, we
compute at the same time and with the same forecasting criteria the factors
and their coefficients.  }

{\normalsize Our formulation is also different because it computes the
forecasting equation for a vector of response time series. Other procedures
concentrates in forecasting a single time series and found the factors and
their coefficients for this single series. In this way the coefficients for
each series and the common components are based on more information,
although we use the restriction that the components will be the same for all
the series. Thus, our approach is expected to be more efficient and powerful
when forecasting many strongly related time series in which the components
are similar for all of them.  }

{\normalsize The general formulation of the core components include as
particular cases many procedures previously proposed, as dynamic principal
components, and dynamic redundancy analysis, and show new light on
procedures as dynamic canonical correlation that can be extended in the
lines described in the paper.  }


 \appendix
\section*{Appendix}
 
\addcontentsline{toc}{section}{Appendix}

 \subsection{Proof of Theorem 1.}
Let  $I_{3}=\{i\in I:|\beta_{i}^{(0)}|>0$ and $g_{i}^{\ast
}(\boldsymbol{\beta}^{(0)})\neq0\}\ $,{ $I_{4}=\{i\in I:\beta_{i}^{(0)}=0$ and
$g_{i}(\boldsymbol{\beta}^{(0)})>\lambda\},$ } and $I_{5}=\{i\in I:\beta
_{i}^{(0)}=0$ and $g_{i}(\boldsymbol{\beta}^{(0)})<-\lambda\}.$
We have that
$I=\cup_{j=1}^{5}I_{j}$,  and, since $\boldsymbol{\beta}^{(0)}$ is not a
critical point, \#$I_{3}+$\#$I_{4}+$\#$I_{5}>0$. Note that from the definition
of $S_{\gamma}$, given $\boldsymbol{\beta}^{(0)}$,  $\beta_{i}^{(1,\eta)}$ is
given by%
\begin{equation}
\beta_{i}^{(1,\eta)}=S_{\eta\lambda}(\beta_{i}^{(0)}-\eta g_{i}%
(\boldsymbol{\beta}^{(0)}))=\left\{
\begin{array}
[c]{ccc}%
\ \beta_{i}^{(0)}-\eta g_{i}^{\ast}(\boldsymbol{\beta}^{(0)})\ \  & \text{if}
& |\beta_{i}^{(0)}-\eta g_{i}(\boldsymbol{\beta}^{(0)})|>\eta\lambda\\
0 & \text{if} & \left\vert \beta_{i}^{(0)}-\eta g_{i}(\boldsymbol{\beta}%
^{(0)})\right\vert \leq\eta\lambda
\end{array}
\right.  .\ \label{m2}%
\end{equation}
 
Since RG$^{\ast}(\boldsymbol{\beta})$ is not\ continuously differentiable at
$\boldsymbol{\beta}$ $\ $if one of its components is $0,$ we are  going to
approximate $\boldsymbol{\beta}^{(0)}\boldsymbol{\ }$by $\boldsymbol{\beta
}^{(0,\delta)}=(\beta_{1}^{(0.\delta)},...,\beta_{p}^{(0.\delta)})^{\prime}$
where:
\begin{equation}
\beta_{i}^{(0.\delta)}=\left\{
\begin{array}
[c]{ccc}%
\delta & \text{if} & i\in I_{1}\\
\beta_{i}^{(0)} & \text{if} & i\in I_{2}\\
\beta_{i}^{(0)} & \text{if} & i\in I_{3}\\
-\delta & \text{if} & i\in I_{4}\\
\delta & \text{if} & i\in I_{5}%
\end{array}
\right.  . \label{m4}%
\end{equation}
We define $\boldsymbol{\beta}^{(1,\eta,\delta)}$ by
\begin{equation}
\boldsymbol{\beta}^{(1,\eta,\delta)}=(\beta_{1}^{(1,\eta,\delta)}%
,...,\beta_{p}^{(1,\eta,\delta)})^{\prime}=\boldsymbol{S}_{\eta\lambda
}(\boldsymbol{\beta}^{(0,\delta)}-\eta g(\boldsymbol{\beta}^{(0,\delta)})).
\end{equation}

Since RG$^{\ast}(\boldsymbol{\beta)}$ is continuously differentiable in a
neighborhood of $\boldsymbol{\beta}^{(0,\delta)}$, by the Mean Value
Theorem  there exist $\eta_{0\text{ }}$ and $\delta_{0}(\eta)$ such that,
\ for $\eta\leq\eta_{0}$ and $\delta\leq\delta_{0}(\eta)$, we have
\begin{equation}
\text{RG}^{\ast}(\boldsymbol{\beta}^{(1,\eta,\delta)})-\text{RG}^{\ast
}(\boldsymbol{\beta}^{(0,\delta)})=\sum_{i=1}^{p}g_{i}^{\ast}%
(\widetilde{\boldsymbol{\beta}}^{(\eta,\delta,\mu(\delta,\eta))})\ (\beta
_{i}^{(1,\eta,\delta)}-\beta_{i}^{(0,\delta)}), %
\end{equation}
where $0<\mu(\eta,\delta)<1$ and by \ (\ref{m2})%
\begin{align}
\widetilde{\boldsymbol{\beta}}^{(\eta,\delta,\mu)}  & =(\widetilde{\beta}%
_{1}^{(\eta,\delta,\mu)},...,\widetilde{\beta}_{p}^{(\eta,\delta,\mu)}%
)=\mu\boldsymbol{\beta}^{(1,\eta,\delta)}+(1-\mu)\ \boldsymbol{\beta
}^{(0,\delta)}\nonumber\\
& =\boldsymbol{\beta}^{(0,\eta,\delta)}-\mu\eta g^{\ast}(\boldsymbol{\beta
}^{(0,\delta)}).\label{tt1d}%
\end{align}

Note that as RG$^{\ast}$ is continuous, $\ $
\begin{align}
\text{RG}^{\ast}(\boldsymbol{\beta}^{(1,\eta)})-\text{RG}^{\ast}%
(\boldsymbol{\beta}^{(0)}) &  =\lim_{\delta\rightarrow0}(\text{RG}^{\ast
}(\boldsymbol{\beta}^{(1,\eta,\delta)})-\text{RG}^{\ast}(\boldsymbol{\beta
}^{(0,\delta)}))\nonumber\\
&  =\sum_{j=1}^{5}\lim_{\delta\rightarrow0}J_{j}^{(\eta,\delta)},\label{tt2}%
\end{align}
where, for $\ 1\leq j\leq5$, %
\begin{equation}
J_{j}^{(\eta,\delta)}=\sum_{i\in I_{j}}g_{i}^{\ast}%
(\widetilde{\boldsymbol{\beta}}_{i}^{(\eta,\delta,\mu(\eta,\delta))}%
)\ (\beta_{i}^{(1,\eta,\delta)}-\beta_{i}^{(0,\delta)}).\label{tt3}%
\end{equation}
Let $\ i$ $\in I_{1}$ then $|g_{i}(\boldsymbol{\beta}^{(0)})|\leq\lambda\ \ $
Compute   first $\beta_{i}^{(1,\eta)}$. From (\ref{m2}) we get
\ $\beta_{i}^{(1,\eta)}=0$ and for the continuity of the two functions, $S$ and
$g_{i}$,  we obtain 
\[
\lim_{\delta\rightarrow0}(\beta_{i}^{(1,\eta,\delta)}-\beta_{i}^{(0,\delta
)})=0,
\]
and, therefore, for all $\eta>0$%
\begin{equation}
\lim_{\delta\rightarrow0}J_{1}^{(\eta,\delta)}=0.\ \label{tt31}%
\end{equation}

The same thing happens with $J_{2}^{(\eta,\delta)}.$ Since $|\beta_{i}%
^{(0)}|>$ $0$ for all $\ i\in I_{2},$ there exists \ \ $\eta_{2}>0$ \ and
$\delta_{2}(\eta)$ $\ $such that for $\eta\leq\eta_{2}$, $\delta\leq\delta
_{2}(\eta)$ and $i\in I_{2}$%

\begin{equation}
\text{sign(}\beta_{i}^{(0,\delta)}-\eta g_{i}(\boldsymbol{\beta}^{(0,\delta
)})=\text{sign(}\beta_{i}^{(0,\delta)})\boldsymbol{\ },  %
\end{equation}
and
\begin{equation}
|\beta_{i}^{(0,\delta)}-\eta g_{i}(\boldsymbol{\beta}^{(0,\delta)}%
)|>\eta\lambda,\text{ }\boldsymbol{\forall}i\in I_{2}. %
\end{equation}

Then, by (\ref{m2}) , for all $i\in I_{2},$ $\eta\leq\eta_{2}$ and $\delta
\leq\delta_{2}(\eta)$
\begin{equation}
\beta_{i}^{(1,\eta,\delta)}\ =\beta_{i}^{(0,\delta)}-\eta g_{i}^{\ast
}(\boldsymbol{\beta}^{(0,\delta)}).\boldsymbol{\ \ } %
\end{equation}

Since for $i\in I_{2}$ we have $g_{i}^{\ast}(\boldsymbol{\beta}_{{}}%
^{(0)})=0\ $ for all $\eta\leq\eta_{2}$,  we have $\lim_{\delta\rightarrow
0}(\beta^{(1,\eta,\delta)}-\beta_{i}^{(0,\delta)})=0$ and, for all
$\eta\leq\eta_{2}$,%
\begin{equation}
\lim_{\delta\rightarrow0}J_{2}^{(\eta,\delta)}=0. \label{tt20}%
\end{equation}

Consider now that $i\in I_{3}$. According to (\ref{tt1d}) we have that  
\begin{equation}
\widetilde{\beta}_{i\ }^{\substack{\\(\eta,\delta,\mu)}}=\beta_{i}%
^{(0,\delta)}-\mu\eta g_{i}(\boldsymbol{\beta}^{(0,\delta)}). %
\end{equation}
Moreover, we can find $\eta_{3}>0$ such that for all $\ \eta\leq\eta_{3}$
there exists $\delta_{3}(\eta)>0$ and $\varepsilon_{3}>0$ satisfying, that for
all $i\in\ I_{3},0<\eta\leq\eta_{3},$ $0<$ $\delta\leq\delta_{3}(\eta)$
and$\ $\ $0<\mu<1,$ we have%
\[
\text{sign(}\beta_{i}^{(0,\delta)})=\text{sign(}\widetilde{\beta}_{i}%
^{\ (\eta,\delta,\mu)}),
\]%
\begin{equation}
|\beta_{i}^{(0,\delta)}-\eta g_{i}^{(0,\delta)}|>\eta\lambda,\text{
}\label{rr24}%
\end{equation}
and%
\[
g_{i}^{\ast}(\boldsymbol{\beta}^{(0,\delta})g^{\ast}%
(\widetilde{\boldsymbol{\beta}}^{\ _{(\eta,\delta,\mu)}})>\varepsilon_{3}.
\]
Besides, by (\ref{m2}) and (\ref{rr24})
\begin{equation}
\beta_{i}^{(1,\eta,\delta)}=S(\beta_{i}^{(0,\delta)}-\eta g_{i}%
(\boldsymbol{\beta}^{(0,\delta}))=\beta_{i}^{(0)}-\eta g_{i}^{\ast
}(\boldsymbol{\beta}^{(0,\delta)})\label{rr14},%
\end{equation}
and taking limit when $\delta\rightarrow0$ we get that, for all $i\in I_{3},$
$0<\eta\leq\eta_{3},$ $0<\mu<1$,
\begin{equation}
\lim_{\delta\rightarrow0}^{{}}g_{i}^{\ast}(\boldsymbol{\beta}^{(0,\delta
)})g_{i}^{\ast}(\widetilde{\boldsymbol{\beta}}_{i}^{\ (\eta,\delta,\mu)}%
)\geq\varepsilon_{3}.\label{rr9}%
\end{equation}

Since by (\ref{rr14})
\begin{equation}
\lim_{\delta\rightarrow0}(\beta_{i}^{(1,\eta,\delta)}-\beta_{i}^{(0,\delta
)})=\ -\lim_{\delta\rightarrow0}\eta g_{i}^{\ast}(\boldsymbol{\beta
}^{(0,\delta)})\ =-\eta g_{i}^{\ast}(\boldsymbol{\beta}^{(0)}),\label{rr7}%
\end{equation}
by (\ref{tt3}) ,(\ref{rr9})\ and (\ref{rr7}) we have that for all $\eta
\leq\eta_{3}$
\begin{equation}
\lim_{\delta\rightarrow0}J_{3,}^{(\eta,\delta)}\leq-\varepsilon_{3}\eta
\#I_{3}.\ \label{tt44}%
\end{equation}

Let $i\in I_{4},$ then since \ $\beta_{i}^{(0,\delta)}=-\delta$ \ $\ \ $and
$g_{i}(\boldsymbol{\beta}^{(0)})>0$ is continuous there exist $\eta_{4}>0,$
$\delta_{4}(\eta)>0$ \ and $\varepsilon_{4}>0$ such that for $\eta\leq\eta
_{4}$ and $\delta\leq$ $\delta_{4}(\eta)$ \
\begin{equation}
\beta_{i}^{(0,\delta)}-\eta g_{i}(\boldsymbol{\beta}^{(0,\delta))}%
)\ <-\eta\lambda, \label{tt15}%
\end{equation}

\begin{equation}
g_{i}(\boldsymbol{\beta}^{(0,\delta)})>\lambda+\varepsilon_{4} \label{tt16}%
\end{equation}

and%
\begin{equation}
\boldsymbol{\ }\beta_{i}^{(0,\delta)}-\eta g_{i}(\boldsymbol{\beta}%
^{(0,\delta)})<-\eta\lambda.\label{tt17}%
\end{equation}

Then, by (\ref{m4}) and (\ref{tt15}) we have%
\[
\beta_{i}^{(1,\eta,\delta)}=\beta_{i}^{(0,\delta)}-\eta(g_{i}%
(\boldsymbol{\beta}^{(0,\delta))})\ -\lambda,
\]
and from (\ref{tt16}) for all $i\in I_{4}$ we obtain .that for all $\eta
\leq\eta_{4}$ and $\delta<\delta_{4}(\eta)\ $%
\[
\beta_{i}^{(1,\eta,\delta)}-\beta_{i}^{(0,\delta)}=-\eta(g_{i}%
(\boldsymbol{\beta}^{(0,\delta)})-\lambda<-\eta\varepsilon_{4}.
\]

Now, for all $\eta\leq\eta_{4}$ and $\delta<\delta_{4}(\eta)$
\begin{equation}
\lim_{\delta\rightarrow0}(\beta_{i}^{(1,\eta_{,}\delta)}-\beta_{i}%
^{(0,\delta)})=-\eta(g_{i}(\boldsymbol{\beta}^{(0)})-\lambda\leq
-\eta\varepsilon_{4}. \label{rr8}%
\end{equation}

From ((\ref{tt1d})) we get that
\[
\widetilde{\boldsymbol{\beta}}^{(\eta,\delta,\mu)}=\boldsymbol{\beta
}^{(0,\delta)}-\mu\eta g(\boldsymbol{\beta}^{(0)}),\
\]
and from (\ref{tt17}) for all $\eta\leq\eta_{4}$ and $\delta<\delta_{4}(\eta)$
and $0\leq\mu\leq1$
\[
\text{ }g_{i}^{\ast}(\ \widetilde{\boldsymbol{\beta}}^{(\eta,\delta,\mu
)})=g_{i}(\boldsymbol{\beta}^{0,\delta}-\mu\eta g(\boldsymbol{\beta
}^{(0,\delta)}))-\lambda>\varepsilon_{4}.
\]
Therefore, for $i\in I_{4},$ $\eta\leq\eta_{4}$, $\delta<\delta_{4}(\eta)\ $
\ \
\begin{equation}
\lim_{\delta\rightarrow0}g_{i}(\ \widetilde{\boldsymbol{\beta}}^{(\eta
,\delta,\mu(\eta,\delta))})>\varepsilon_{4},\label{rr10}%
\end{equation}
and by (\ref{tt3}), (\ref{rr8} and (\ref{rr10}) for all $\eta\leq
\eta_{4}$
\begin{equation}
\lim_{\delta\rightarrow0}J_{4}^{(\eta,\delta)}\leq-\eta\varepsilon_{4}%
^{2}\#I_{\substack{4\\}}.\ \ \label{tt18}%
\end{equation}

Similarly as we have proved that there exists $\eta$ \ such that $\lim
_{\delta\rightarrow0}J_{4}^{(\eta,\delta)},<0$ it may be proved that there
exists $\eta_{5}$ such that for $\eta\leq\eta_{5}$ $\ $%
\begin{equation}
\lim_{\delta\rightarrow0}J_{5}^{(\eta,\delta)}\leq-\eta_{5}\#I_{\substack{5\\}%
}.\label{tt7}%
\end{equation}

Define\ $\ \eta_{{}}^{\ast}=\min(\eta_{0},\min_{2\leq j\leq5,}\eta_{j})$ ,
then \ \ by, (\ref{tt2}), (\ref{tt3}), (\ref{tt31}), (\ref{tt20}),
(\ref{tt44}), \ (\ref{tt18}), (\ref{tt7}) and $\#I_{3}+\#I_{4}+\#I_{5}>0$
imply that
\[
RG^{\ast}(\boldsymbol{\beta}^{(1,\eta^{\ast})})<RG^{\ast}(\boldsymbol{\beta
}^{(0)}),\ \
\]
and the Theorem is proved.

\subsection{Proof of Theorem 2.}

Since the sequence RG$^{\ast}(\boldsymbol{\beta}^{(j)})$ is decreasing and
bounded below by $0,$ there exists
\[
t_{0}=\lim_{j\rightarrow\infty}\text{RG}^{\ast}(\boldsymbol{\beta}^{(j)}),
\]
and since RG$^{\ast}$ is continous, we have
\[
\text{RG}^{\ast}(\boldsymbol{\beta}_{0})=t_{0}.
\]

\ Suppose that $\boldsymbol{\beta}_{0}$ is not a critical point. Then, by
theorem 1,\ there exists $\eta_{0}\ $such that if $\boldsymbol{\beta}%
_{1}=\mathbf{S}_{\lambda\eta_{0}}(\boldsymbol{\beta}_{0}-\eta_{0}%
g(\boldsymbol{\beta}_{0})),$ then RG$^{\ast}(\boldsymbol{\beta}_{1}%
)=t_{1}<t_{0}$ and
\[
t_{1}=\lim_{k\rightarrow\infty\ \ }\text{RG}^{\ast}\text{(}\mathbf{S}%
_{\lambda\eta_{0}}(\boldsymbol{\beta}_{{}}^{(j_{k})}-\eta_{0}%
g(\boldsymbol{\beta}_{{}}^{(j_{k})}))).
\]
By the continuity \ of RG$^{\ast}$ \ and $\mathbf{S}$ there exist $k_{0}$ such
that%
\[
\text{RG}^{\ast}(\mathbf{S}_{\lambda\eta_{0}}(\boldsymbol{\beta}_{{}%
}^{(j_{k_{0}})}-\eta_{0}g(\boldsymbol{\beta}_{{}}^{(j_{k_{0}})})))<t_{0}.
\]
However, the selected value $\eta_{j_{k_{0}}\text{ }}$ for the $j_{k_{0}}$
step of the algorithm yields%
\[
\text{RG}^{\ast}(\mathbf{S}_{\lambda\eta_{j_{k_{0}}}}(\boldsymbol{\beta}_{{}%
}^{(j_{k_{0}})}-\eta_{0}g(\boldsymbol{\beta}_{{}}^{(j_{k_{0}})})))=\text{RG}%
^{\ast}(\boldsymbol{\beta}_{{}}^{(j_{k_{0}}+1)})>t_{0},
\]
and this contradicts the definition of $\eta_{j_{k_{0}}}$ as
\[
\eta_{j_{k_{0}}}=\arg\min_{\eta}\text{RG}^{\ast}(\mathbf{S}_{\lambda\eta
}(\boldsymbol{\beta}_{{}}^{(j_{k_{0}})}-\eta g(\boldsymbol{\beta}_{{}%
}^{(j_{k_{0}})})).
\]
This proves the theorem.

 \subsection{Proof of Theorem 3.}

  Let $\mathbf{u\in }R^{q},$ then $\partial \mathbf{||\mathbf{u}||%
}^{2}/\partial \mathbf{u}=2\mathbf{u}$. Then 
\begin{equation}
\frac{\partial G_{1}(\boldsymbol{\boldsymbol{\beta ,\Gamma }})}{\partial 
\boldsymbol{\boldsymbol{\beta }}}=-\frac{1}{(T-h-k)}\sum_{t=k+1}^{T-h}%
\mathcal{X}_{t}^{\prime }\boldsymbol{\ }\mathbf{\Gamma }^{\prime }({\mathbf{y%
}_{t+h}-\boldsymbol{\mathbf{\Gamma }}}\ {\mathcal{X}_{t}\boldsymbol{\beta }\
)\mathbf{\ }=0}, \label{Dbeta}
\end{equation}%
 then, setting ${\normalsize \partial G_{1}(%
\boldsymbol{\boldsymbol{\beta ,\Gamma }})/}\partial \boldsymbol{\boldsymbol{%
\beta }}${\normalsize \ }$=\mathbf{0}${\normalsize \ and solving for $%
\boldsymbol{\beta }_{{}}$ we get (\ref{eq1}). }

{\normalsize Equation (\ref{eq20}) is the least squared solution to explain
the vector $\mathbf{y}_{t}$ with the vector $\mathbf{f}_{t}$. }
 
\subsection{Proof of Theorem 4} 

{\normalsize Let 
\begin{equation*}
\mathbf{\Sigma }(\boldsymbol{\beta },\mathbf{\Gamma })\mathbf{=}\frac{1}{%
(T-h-k)}\sum_{t=k+1}^{T-h}\mathbf{e}_{t+h|t}(\boldsymbol{\beta }\ ,\mathbf{%
\Gamma })\mathbf{e}_{t+h|t}^{\prime }(\boldsymbol{\beta },\mathbf{\Gamma })%
\mathbf{,}
\end{equation*}%
\ where $\mathbf{e}_{t+h|t}(\boldsymbol{\beta },\mathbf{\Gamma })\mathbf{=y}%
_{t+h}\mathbf{-C}_{t}\boldsymbol{\beta ,}$ and $\mathbf{C}_{t}\mathbf{%
=\Gamma }\mathcal{X}_{t}.$ We have to find $\boldsymbol{\beta }_{{}}$ such
that%
\begin{equation}
\frac{\mathbf{\partial }G_{2}(\boldsymbol{\beta },\mathbf{\Gamma })}{%
\partial \boldsymbol{\beta }}=\frac{\mathbf{\partial |\Sigma }(%
\boldsymbol{\beta },\mathbf{\Gamma })\mathbf{|}}{\mathbf{\partial }%
\boldsymbol{\beta }}\boldsymbol{=0.}  \label{ig00}
\end{equation}%
We start getting an expression for 
\begin{equation*}
\mathbf{D}_{k}\mathbf{=}\frac{\mathbf{\partial \Sigma }(\boldsymbol{\beta },%
\mathbf{\Gamma })}{\mathbf{\partial }\beta _{k}}\mathbf{\ =}\frac{1}{(T-h-k)}%
\sum_{t=k+1}^{T-h}\mathbf{(\partial (e}_{t+h|t}(\boldsymbol{\beta },\mathbf{%
\Gamma })\mathbf{e}_{t+h|t}^{\prime }(\boldsymbol{\beta },\mathbf{\Gamma }))%
\mathbf{/\partial }\mathbb{\beta }_{k}\mathbf{),}
\end{equation*}%
where $\mathbb{\beta }_{k}$ is the $k$-th component of the vector $%
\boldsymbol{\beta }_{{}}$. }

{\normalsize Since%
\begin{equation*}
\mathbf{\partial (e}_{t+h|t}(\boldsymbol{\beta },\mathbf{\Gamma }_{{}})%
\mathbf{e}_{t+h|t}^{\prime }(\boldsymbol{\beta },\mathbf{\Gamma }))\mathbf{/%
\mathbf{\partial }}\beta _{k}\mathbf{=-c}^{(k)}\mathbf{e}_{t+h|t}^{\prime }(%
\boldsymbol{\beta },\mathbf{\Gamma })\mathbf{-e}_{t+h|t}(\boldsymbol{\beta },%
\mathbf{\Gamma })\mathbf{c}^{(k)\prime },
\end{equation*}%
where $\mathbf{c}^{(k)}$ is the $k-$th column of $\mathbf{C}_{t}$, then, 
\begin{equation}
\mathbf{D}_{k}\mathbf{=-}\frac{1}{(T-h-k)}\sum_{t=k+1}^{T-h}(\mathbf{c}^{(k)}%
\mathbf{e}_{t+h|t}^{\prime }(\boldsymbol{\beta },\mathbf{\Gamma })\mathbf{-e}%
_{t+h|t}(\boldsymbol{\beta },\mathbf{\Gamma })\mathbf{c}^{(k)\prime }).
\label{igg}
\end{equation}%
Given a square matrix $\mathbf{A(}x\mathbf{),}$ denoting its trace by tr$(%
\mathbf{A(}x\mathbf{))}$ where $x$ is a scalar variable, then, by formula
(46) of \cite{petersen} $\mathbf{\partial |A(}x\mathbf{)}\mathbf{|/\partial }%
x=\mathbf{|A(}x\mathbf{)|}{\text{tr}}(\mathbf{A(}x)\mathbf{\partial
A/\partial }x\mathbf{)}^{-1}$. Therefore, we have%
\begin{equation}
\mathbf{\partial |\Sigma }(\boldsymbol{\beta },\mathbf{\Gamma })\mathbf{%
|/\partial }\beta _{k}\mathbf{=|\Sigma }(\boldsymbol{\beta },\mathbf{%
\Gamma })\mathbf{|(}{\text{tr}}\mathbf{(\Sigma }(\boldsymbol{\beta },%
\mathbf{\Gamma })^{-1}\mathbf{D}_{k}\mathbf{).}  \label{ig1}
\end{equation}%
Since tr$(\mathbf{ab}^{\prime })=\mathbf{a}^{\prime }\mathbf{b,}$ by (\ref%
{igg}) we get%
\begin{align}
{\text{tr}}(\mathbf{\Sigma }^{-1}\mathbf{D}_{k})& =-\frac{1}{(T-h-k)}%
\sum_{t=k+1}^{T-h}\left[ ({\text{tr}}(\mathbf{\Sigma }^{-1}\mathbf{c}%
^{(k)}\mathbf{e}_{t+h|t}^{\prime }))-({\text{tr}}(\mathbf{\Sigma }^{-1}%
\mathbf{e}_{t+h|t}\mathbf{c}^{(k)\prime }))\right]  \notag \\
& \mathbf{=}-\frac{1}{(T-h-k)}\sum_{t=k+1}^{T-h}\mathbf{((e}_{t+h|t}^{\prime
}\mathbf{\Sigma }^{-1}\mathbf{c}^{(k)}\mathbf{)-(c}^{(k)\prime }\mathbf{%
\Sigma }^{-1}\mathbf{e}_{t+h|t}\mathbf{))}  \notag \\
& \mathbf{=}-\frac{2}{(T-h-k)}\sum_{t=k+1}^{T-h}\mathbf{(c}^{(k)\prime }%
\mathbf{\Sigma }^{-1}\mathbf{e}_{t+h|t}\mathbf{).}  \label{ig2}
\end{align}%
Equations (\ref{igg}), (\ref{ig1}) and (\ref{ig2}) imply 
\begin{equation*}
\ \frac{\mathbf{\partial |\Sigma }(\boldsymbol{\beta },\mathbf{\Gamma })%
\mathbf{|}}{\mathbf{\partial }\beta _{k}}\mathbf{\ }=\mathbf{-}\frac{2|%
\mathbf{\Sigma }\mathbf{|}}{(T-h-k)}\sum_{t=k+1}^{T-h}\mathbf{(c}%
^{(k)\prime }\mathbf{\Sigma }^{-1}\mathbf{e}_{t+h|t}\mathbf{)}
\end{equation*}%
and we conclude that 
\begin{equation}
\frac{\mathbf{\partial }G_{2}(\boldsymbol{\beta },\mathbf{\Gamma })}{%
\partial \boldsymbol{\beta }}=\frac{\mathbf{\partial |\Sigma }(%
\boldsymbol{\beta },\mathbf{\Gamma })\mathbf{|}}{\mathbf{\partial }%
\boldsymbol{\beta }}\mathbf{=}-\frac{2|\mathbf{\Sigma }\mathbf{|}}{%
(T-h-k)}\sum_{t=k+1}^{T-h}\mathbf{(C}_{t}^{\prime }\mathbf{\Sigma }^{-1}%
\mathbf{e}_{t+h|t}\mathbf{)}.  \label{DG2}
\end{equation}%
Assuming that $\mathbf{|\Sigma }\mathbf{|>0,}$\textbf{\ }(\ref{ig00})%
\textbf{\ }is equivalent to%
\begin{equation*}
\frac{1}{(T-h-k)}\sum_{t=k+1}^{T-h}(\mathbf{C}_{t}^{\prime }\mathbf{\Sigma }%
^{-1}\mathbf{(y}_{t+h}\mathbf{-C}_{t}\boldsymbol{\beta }\mathbf{)})=%
\mathbf{0.}
\end{equation*}
\ \ \ \ Then, solving for $\boldsymbol{\beta }_{{}}$, we get (\ref{eq3})$.$ }

{\normalsize To prove (\ref{TE22}) note that if $\widehat{\boldsymbol{\beta }}%
$ is known,   
\begin{equation*}
\widehat{\mathbf{\Gamma }}=\arg \min_{\mathbf{\Gamma }}\left\vert
\sum_{t=k+1}^{T-h}(\mathbf{y}_{t+h}-\mathbf{\Gamma }\mathbf{f}_{t}(%
\widehat{\boldsymbol{\beta }}\boldsymbol{))}(\mathbf{y}_{t+h}-\mathbf{%
\Gamma } \mathbf{f}_{t}(\widehat{\boldsymbol{\beta }} \boldsymbol{))}%
^{\prime }\right\vert ,
\end{equation*}%
but this is the equation of the maximum likelihood estimator for the normal
multivariate model where $Y$ is the outcome matrix and $F(\widehat{%
\boldsymbol{\beta }})$ is the regressor matrix. It is well known that
the solution to this problem is given by (\ref{TE22}), that is, by
coordinate-wise least squares estimators.}

\bibliographystyle{authordate1}
\bibliography{mtsred}	

\end{document}